\documentclass[%
 reprint,
 amsmath,amssymb,
 aps,
 prx,
]{revtex4-2}

\usepackage[normalem]{ulem}
\usepackage{graphicx}
\usepackage{subcaption}
\usepackage{dcolumn}
\usepackage{bm}
\usepackage{xcolor}
\usepackage{caption}
\usepackage{ragged2e}
\usepackage[colorlinks=true]{hyperref}
\usepackage{cleveref}
\crefname{figure}{Fig.}{Figs.}
\crefname{equation}{Eq.}{Eqs.}
\crefname{section}{Sec.}{Secs.}
\crefname{subsection}{Sec.}{Secs.}
\crefname{appendix}{App.}{Apps.}

\newcommand{\timederiv}{\mathrm{d}_t}
\newcommand{\superopbr}[1]{#1}

\newcommand{\bra}[1]{{\left\langle  #1 \right|}}
\newcommand{\ket}[1]{{\left|  #1 \right\rangle}}

\begin{document}

\preprint{APS/123-QED}

\title{Quantum statistical enhancement of collective behaviour in a bosonic active Ising model}

\author{Kian L.~Assent}
\email{assent@tu-berlin.de}
\author{Emil Strauch}
\author{Sabine H.~L.~Klapp}
\author{André Eckardt}
\author{Alexander Schnell}
\email{schnell@tu-berlin.de}
\affiliation{%
 Technische Universität Berlin, Institut für Physik und Astronomie, Berlin, Germany
 }%

\date{\today}

\begin{abstract}
Collective behaviour such as flocking (the collective motion of a spontaneously formed group along a common direction) or aster formation (the binding of opposing flocks, inhibiting each others motion)  are  intriguing emergent phenomena in active systems with local alignment rules. 
Until recently, their occurrence was mainly studied for classical systems, a prime example being the active Ising model (AIM), which translates the main ingredients of flocking and aster formation (i.e., alignment and self-propulsion) to a lattice framework.
Here we introduce and study a one-dimensional (1D) quantum lattice variant of the AIM, based on ideal bosons with a spin degree of freedom.
We find that both the collective behaviours of the 1D classical model, flocking and aster formation, are markedly enhanced by the bosonic quantum statistics. 
This contrasts with a recent quantum generalization of the AIM based onto hard-core bosons [Khasseh et al., Phys.~Rev.~Lett.\  {\bf135}, 248302 (2025)], where flocking, but neither its quantum-statistical stabilization nor aster states were observed as a consequence of interactions. 
Moreover, we investigate the competition of this quantum statistical stabilization of collective phases with their suppression by the quantum fluctuations induced by a transverse external magnetic field.
\end{abstract}

\maketitle


\section{Introduction}

Over the last decades, we saw great progress in the engineering and control of quantum matter \cite{Bloch2012,Carusotto2013,qm_engineer_1,Daley2022, qm_engineer_3,qm_engineer_4,qm_engineer_6}. 
Here an important class of models describes driven-dissipative systems \cite{Sieberer2016,Bloch2022,Sieberer2025a,vorberg_nonequilibrium_2015,Leymann2017a}, which are inherently out-of-equilibrium  and are described by  the theory of open quantum systems 
\cite{OQS,OQS_1,OQS_2,OQS_3,OQS_5}. A prominent example are exciton-polariton condensates \cite{Carusotto2013,Bloch2022} where a steady state emerges by laser pumping of the photonic component acting against the inherent losses in the system. In these systems, recently nonequilibrium universality of the Kardar–Parisi–Zhang (KPZ) class has been observed \cite{Fontaine2022a}.
Also, there have been experimental realizations of condensates of light \cite{Klaers2010a,Bloch2022,Vretenar2021} by optically pumping dye-filled cavities.
Another example are superconducting circuits \cite{Krantz2019,Blais2021}, where nonlinearities have been used to engineer (driven-dissipative) artificial atoms \cite{Blais2021}. Here, both Mott insulator states \cite{Ma2019} and (fractional) quantum Hall states \cite{Rosen2024,Clark2020,Wang2024b} have been realized.
Theoretical works on driven-dissipative systems include general theories for condensation in non-equilibrium steady states \cite{Vorberg2013,vorberg_nonequilibrium_2015,Schnell2017,schnell_number_2018}, universality in driven open matter \cite{Altman2015,Zelle2024,Sieberer2025a}, bath-engineering of effective low-entropy many-body states in atomic- and circuit QED platforms \cite{Iadecola2015a,Shirai2015a,Shirai2016a,Schnell2024a,Petiziol2022}, and induced long-range order by driving a system out of equilibrium \cite{Schnell2017,Schnell2023,Gladilin2022}.

In parallel, there has been tremendous progress in understanding the physics of classical active systems consisting of motile ``agents" (such as bacteria, self-propelled colloids, or birds) \cite{Marchetti2013,TeVrugt2025,Bechinger2016}. In these systems, individual agents or particles are fueled by internal energy sources or the environment, leading to persistent motion on the particle level and a range of intriguing collective many-body behaviours. Paradigmatic examples are flocking \cite{vicsek, Solon2015,Chate2020} (spontaneous collective motion induced by alignment, e.g., of birds), formation of ordered bands and clusters \cite{Solon2015}, asters \cite{classical_flocking}, motility-induced phase separation \cite{Cates2015,Bialke2013} (in purely repulsive active systems) and mesoscale turbulence \cite{bacteria2,ising_4} (a spatio-temporal pattern formation phenomenon accompanied by large-scale vortices).

In view of these developments, it seems natural to ask whether and how the collective behaviours seen in classical active matter can be observed and modified in quantum systems \cite{quantum_flocking,qm_engineer_2,Penner2025,Steiner2026,Burgardt2026}.
Recently, an important step in this direction has been made by Khasseh et al.~\cite{quantum_flocking}, who proposed a 1D dissipative  model based on hard-core bosons for an active quantum matter system exhibiting flocking as a result of incoherent processes describing spin alignment and spin-dependent biased diffusion.  In this system, they also studied the suppression of collective behaviour by quantum fluctuations introduced by a transverse magnetic field that counteracts collective spin alignment.  

In the present paper, we focus as well on quantum flocking in one dimension, but consider ideal bosons without mutual repulsion. In this way, we are even closer to the original classical models of flocking, particularly the active Ising model (AIM) \cite{Solon2013,ising3,classical_flocking}, a minimalistic spin lattice model involving only alignment and biased diffusion. In this model, we do not only find flocking, but as a second collective behaviour also aster states, as they were previously found in classical models \cite{Solon2013}. 
Moreover, we also find that, different from the case of hard-core bosons, the bosonic quantum statistics markedly enhances collective behaviour for filling factors $\ge \mathcal{O}(1)$. Namely, in contrast to the suppression of collective alignment by the quantum fluctuations induced of a transverse field \cite{quantum_flocking}, which we equally find in our model, bosonic indistinguishability constitutes a second quantum effect that strengthens local alignment by the tendency of bosons to occupy the same state. 


Concretely, we study ideal bosons with two relevant spin states on a one-dimensional lattice. 
We translate the classical master equation which governs the classical active Ising model to a quantum master equation by introducing the corresponding Lindblad jump operators.
These include a spin-dependent biased diffusion as well as a collective Ising-type spin-alignment process. Additionally, we also include a Hamiltonian describing the impact of an external magnetic field transverse to the quantization axis of the spin alignment. Even though the Hamiltonian is quadratic with respect to the bosonic annihilation and creation operators and, thus, describes a non-interacting system, both dissipative terms enter at least as quartic terms in the master equation and, therefore, correspond to interactions.   
In order to simulate the dynamics of our model, we derive equations of motion for the single-particle density matrix and perform a mean-field approximation to find kinetic equations, which govern the system dynamics in presence of the dissipator. 


This paper is organized as follows: In \cref{aim_intro} we recapitulate the classical active Ising model and its phases by performing classical Monte-Carlo simulations. \cref{quantum_master_eq} generalizes the classical active Ising model to ideal bosons by introducing corresponding  Lindblad jump operators. We discuss how the model can be solved for large system sizes by driving kinetic equations of motion via a mean-field approximation. Then, in \cref{simulations}, we discuss the results of our simulations, namely the existence of the quantum flocking- and aster phases. We demonstrate their enhancement over the classical phases due to the bosonic quantum statistics as well as their suppression by quantum fluctuations induced by a transverse magnetic field. Additionally, we derive analytical estimates for the phase boundaries in the quantum model. Lastly, in \cref{sec:conclusions}, we conclude our findings and discuss remaining open questions.

\section{Classical active Ising model}\label{aim_intro}

\subsection{Definition of the model}
Before introducing our active quantum model, let us recapitulate the classcial active Ising model.

The active Ising model (AIM) \cite{Solon2013,ising3,classical_flocking} is a prominent classical lattice model that involves the two crucial ingredients required for flocking: self-propulsion, directed by an internal classical spin-1/2 degree of freedom with two relevant states, and ferromagnetic alignment, while no excluded volume or other interactions are present. 
It has been originally introduced as a lattice variant of the famous Vicsek model \cite{vicsek} of flocking, an off-lattice system of self-propelled point particle with polar (XY-type) interactions.


\begin{figure}[t]
  \centering
  \begin{subfigure}[t]{0.23\textwidth}
    \centering
    \caption{\raggedright}
    \includegraphics[width=\textwidth]{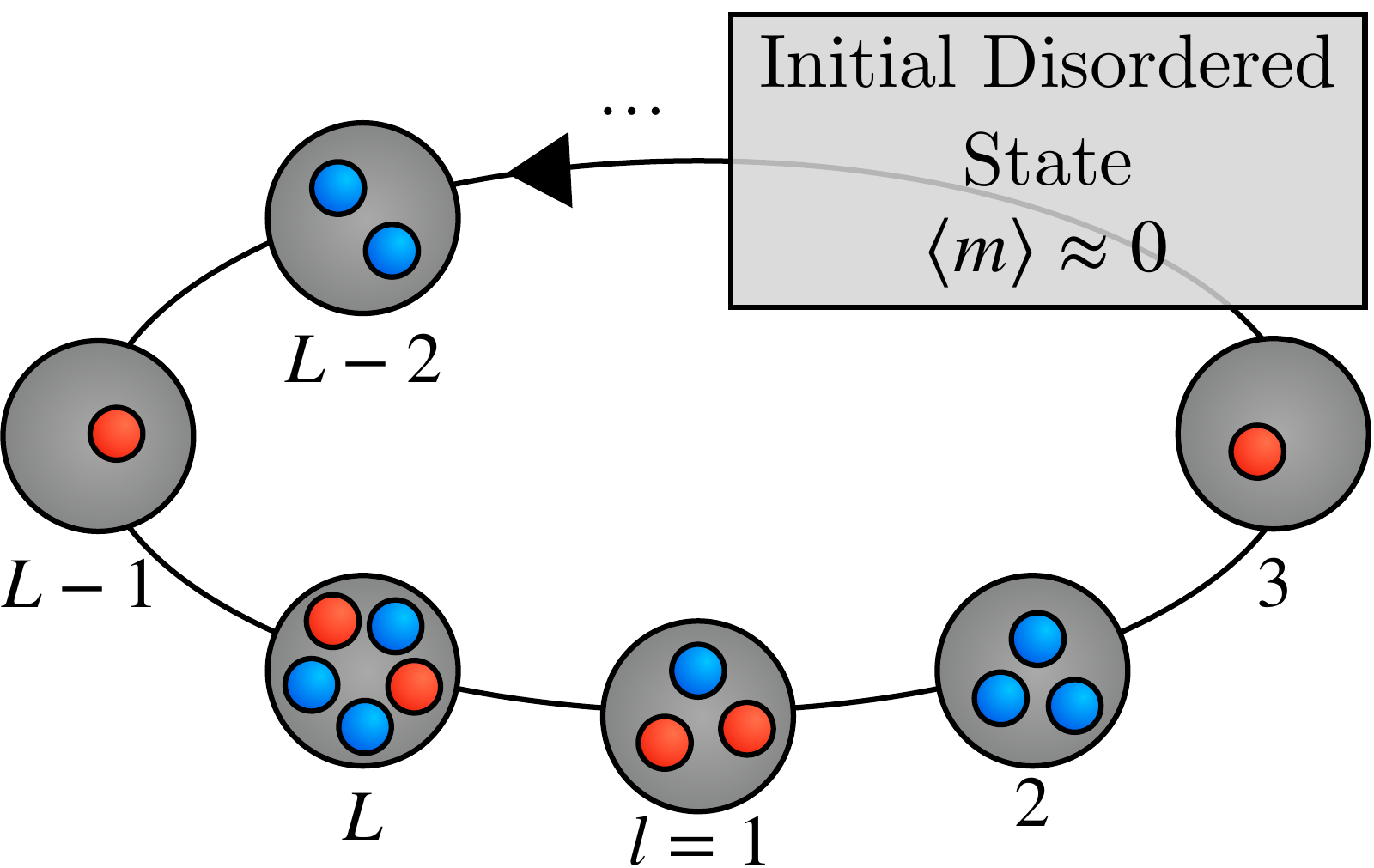}
    \label{aim:b}
  \end{subfigure}\hfill
    \begin{subfigure}[t]{0.23\textwidth}
    \centering
    \caption{\raggedright}
    \includegraphics[width=\textwidth]{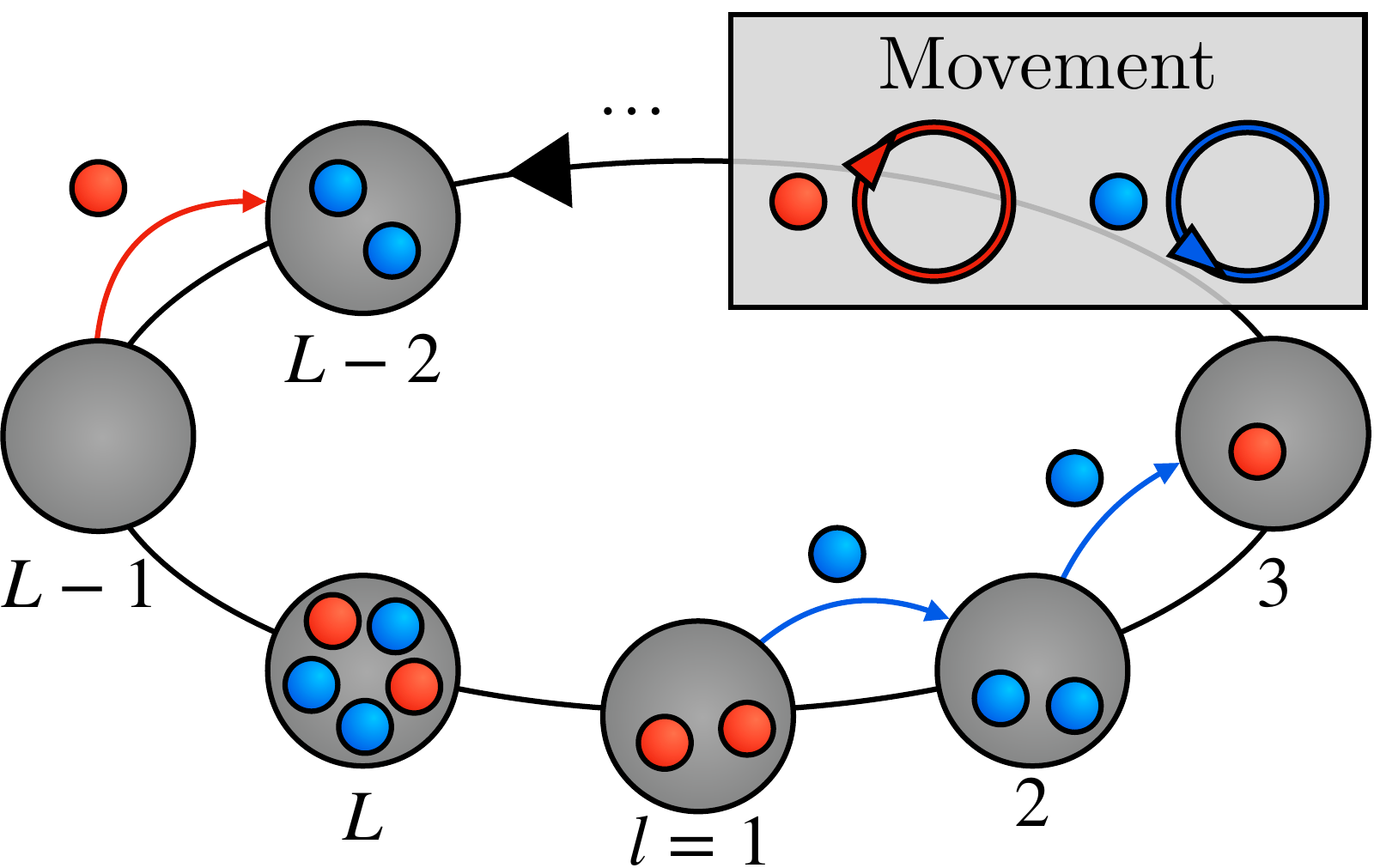}
    \label{aim:a}
  \end{subfigure}

  \vspace{-2.5em}
 
  \begin{subfigure}[t]{0.23\textwidth}
    \centering
    \caption{\raggedright}
    \includegraphics[width=\textwidth]{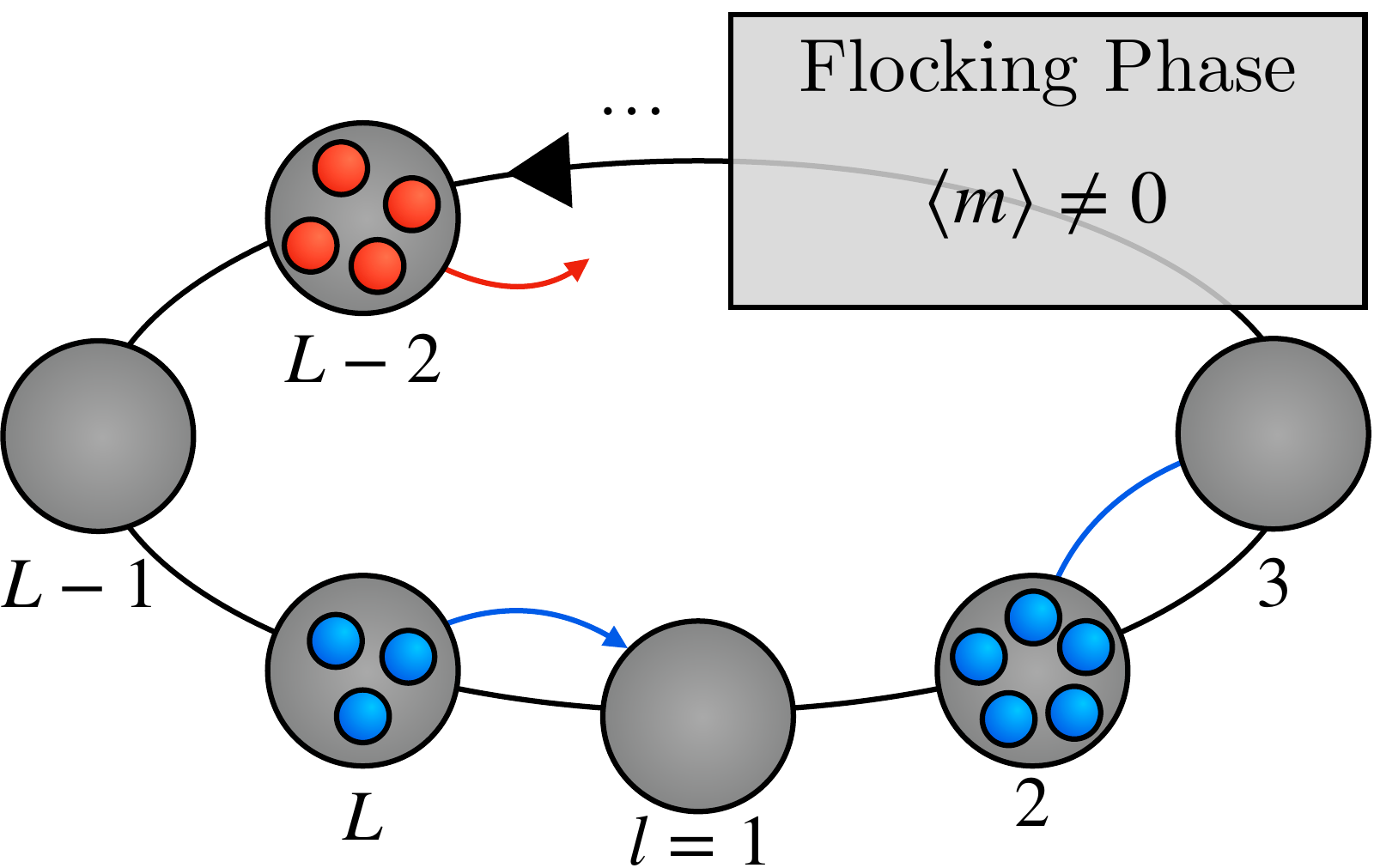}
    \label{aim:c}
  \end{subfigure}\hfill
    \begin{subfigure}[t]{0.23\textwidth}
    \centering
    \caption{\raggedright}
    \includegraphics[width=\textwidth]{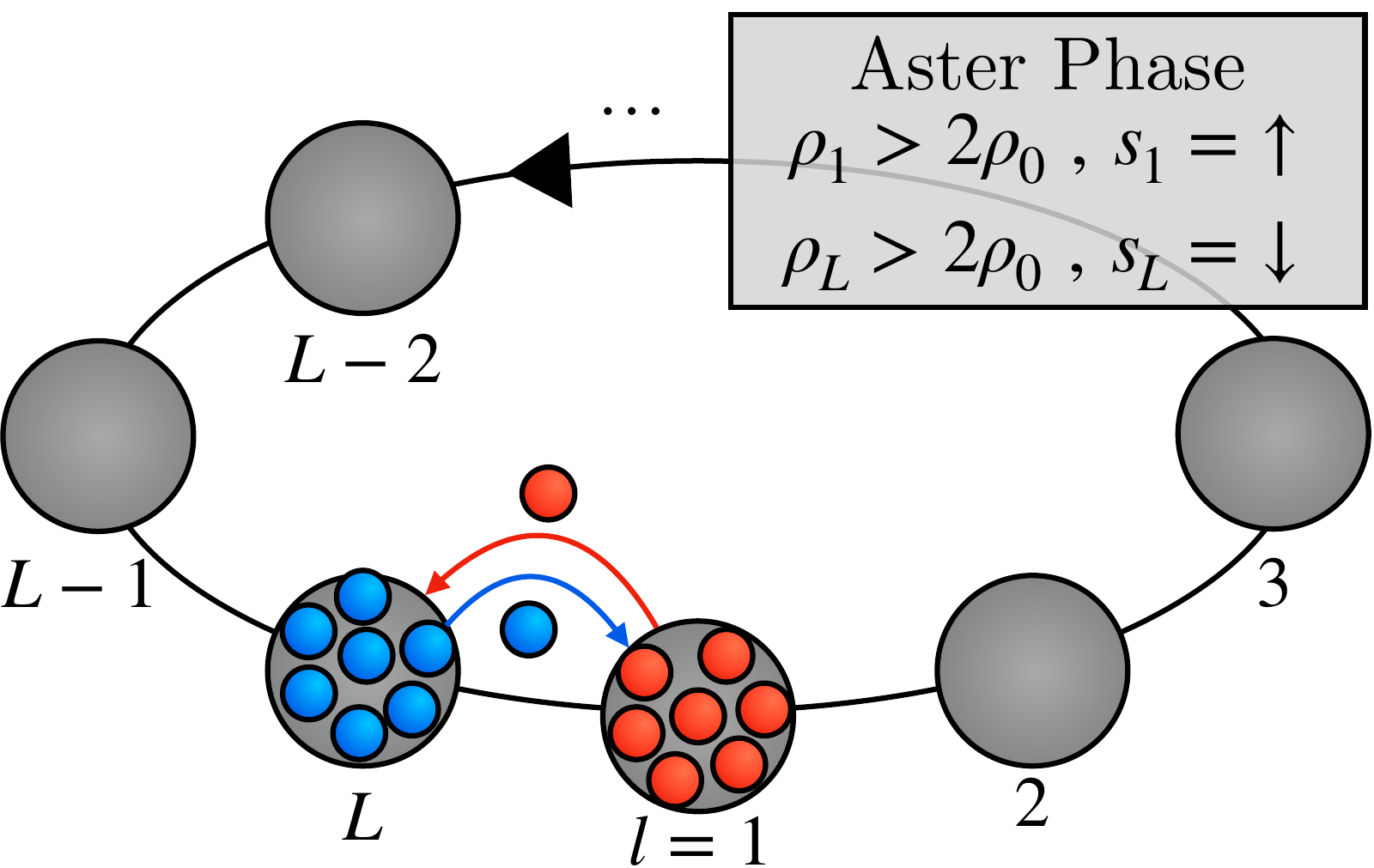}
    \label{aim:d}
  \end{subfigure}
  \vspace{-1.5em}
  \caption{\label{aim:ges}\justifying 
  (a) The particle's spin species governs its preferred direction of movement. (b) The lattice sites of the AIM are initially occupied by a random distribution of particles.
  (c) Large groups of same-species particles moving around the AIM are considered a flock, while (d) a large group of opposite-species-particles remaining dormant on neighboring sites are called an aster.}
  \vspace{-1.5em}
\end{figure}

In this work, we focus on a one-dimensional AIM.
It consists of a ring of lattice sites $(l=1,\dots,L)$ with periodic boundary conditions (i.e.~we identify $l=L+1$ with $l=1$). We use $N$ particles with spin species $s=\uparrow$ and $\downarrow$ (marked by red and blue, respectively, in \cref{aim:b}) that populate the lattice. 
For each lattice site $l$, the number of particles pointing up or down is denoted by $n_{l\uparrow}$ and $n_{l\downarrow}$, respectively. Importantly, there is no repulsion between the particles, therefore the occupation numbers $n_{l\uparrow(\downarrow)}$ can take any value between zero and $N$.
The particles are able to hop from one site to another, with a preferred  direction of hopping depending on their spin (\cref{aim:b}). 
In the AIM, this enters via the rate for diffusion of a particle to the right (from $l$ to $l+1$)
\begin{align}
    W_{\mathrm{H,r}} = D(1 + s \varepsilon),
    \label{rate-diffusion-right}
\end{align}
where we identify the integers $s=1$ with spin $s=\uparrow$ and $s=-1$ with $s=\downarrow$. Analogously, for the diffusion to the left we assume
\begin{align}
    W_{\mathrm{H,l}} = D(1 - s \varepsilon),
    \label{rate-diffusion-left}
\end{align}
leading to biased diffusion of the $s=\uparrow$ to the right and $s=\downarrow$ to the left. The strength of this bias is controlled by the   activity parameter $\varepsilon \in [0, 1]$.
Additionally, a particle at site $l$ can flip its spin state from $s$ to $-s$ at rate 
\begin{align}
     W_\mathrm{F}=\gamma e^{-\beta s \frac{n_{l\uparrow}-n_{l\downarrow}}{n_{l\uparrow}+n_{l\downarrow}}}
     =\gamma e^{-\beta s \frac{m_l}{\rho_l}}
    \label{rate-flip}
\end{align}
where $\beta=1/T$ is the inverse temperature, which, like the temperature $T$, is dimensionless, as we set the Boltzmann constant to one and express all energies in terms of the amplitude of the spin-spin interaction. 
In Eq.~(\ref{rate-flip}), we have introduced the (unbounded) local density $\rho_l=n_{l\uparrow}+n_{l\downarrow}$ and the local magnetization $m_l=n_{l\uparrow}-n_{l\downarrow}$, with
$-\rho_l\leq m_l\leq \rho_l$. 
Equation~(\ref{rate-flip}) reflects that flipping the spin \textit{against} the majority of spin directions on the same lattice site is highly disfavoured relative to the opposite flipping process. Combined with the biased diffusion, this on-site alignment coupling leads to the emergence of both flocking and aster formation \cite{Solon2013,classical_flocking}.

The spin-flip rate (\ref{rate-flip}) can be derived from a Hamiltonian 
of the form $\sum_l m_l^2/\rho_l$ (which is dimensionless, as its amplitude serves as the unit of energy, as mentioned already above) describing a fully connected Ising model on
each lattice site. In this sense, $W_\mathrm{F}$ fulfills detailed
balance (in the absence of diffusion). Here, fully connected refers to the fact that all the spins gathering on one lattice site interact with each other.
Note that other definitions of the rate lead to similar flocking behavior \cite{scandolo_active_2023} showing that the definition of $W_\mathrm{F}$ in this fully out-of-equilibrium model is, to some extent, arbitrary (as long as it favors alignment).

For later purposes, here we also present the classical master equation corresponding to the processes defined above.
Denoting the state of the entire system as $\mathbf{n} = (n_{1 \uparrow}, n_{1 \downarrow},n_{2 \uparrow}, n_{2 \downarrow} \dots)$ we find
\begin{align}
\begin{split}
\timederiv p_\mathbf{n} = \sum_{l, s}& \Bigl\{ D(1+s\varepsilon) \left[(n_{l-1s}+1) p_{\mathbf{n}_{l-1s}}- n_{ls} p_{\mathbf{n}}\right] \\
&+ D(1-s\varepsilon) \left[(n_{l+1s}+1) p_{\mathbf{n}_{l+1s}}- n_{ls} p_{\mathbf{n}}\right]\\
& + \gamma  e^{\frac{\beta}{\rho_l}  (n_{ls}-n_{l,-s}-2)}(n_{l,-s}+1) p_{\mathbf{n}_{l,-s}} \\
&-  \gamma  e^{-\frac{\beta}{\rho_l}  (n_{ls}-n_{l,-s})} n_{ls} p_{\mathbf{n}}\Bigr\}.
\end{split}
\label{master-classical}
\end{align}
Here, we have defined the configuration $\mathbf{n}_{l\pm1s} = (\dots,  n_{ls}-1, n_{l\pm1s}+1,\dots)$ which results from the configuration $\mathbf{n}$ by moving one spin-$s$ particle on site $l$ to the right or left. Similarly, we define $\mathbf{n}_{l, -s} = (\dots,  n_{ls}-1, n_{l,-s}+1,\dots)$ which results from flipping the spin of one particle on site $l$ from $s$ to $-s$.

\begin{figure}[b]
  \centering
  \begin{subfigure}[t]{0.24\textwidth}
    \centering
    \caption{\raggedright}
    \includegraphics[width=\textwidth]{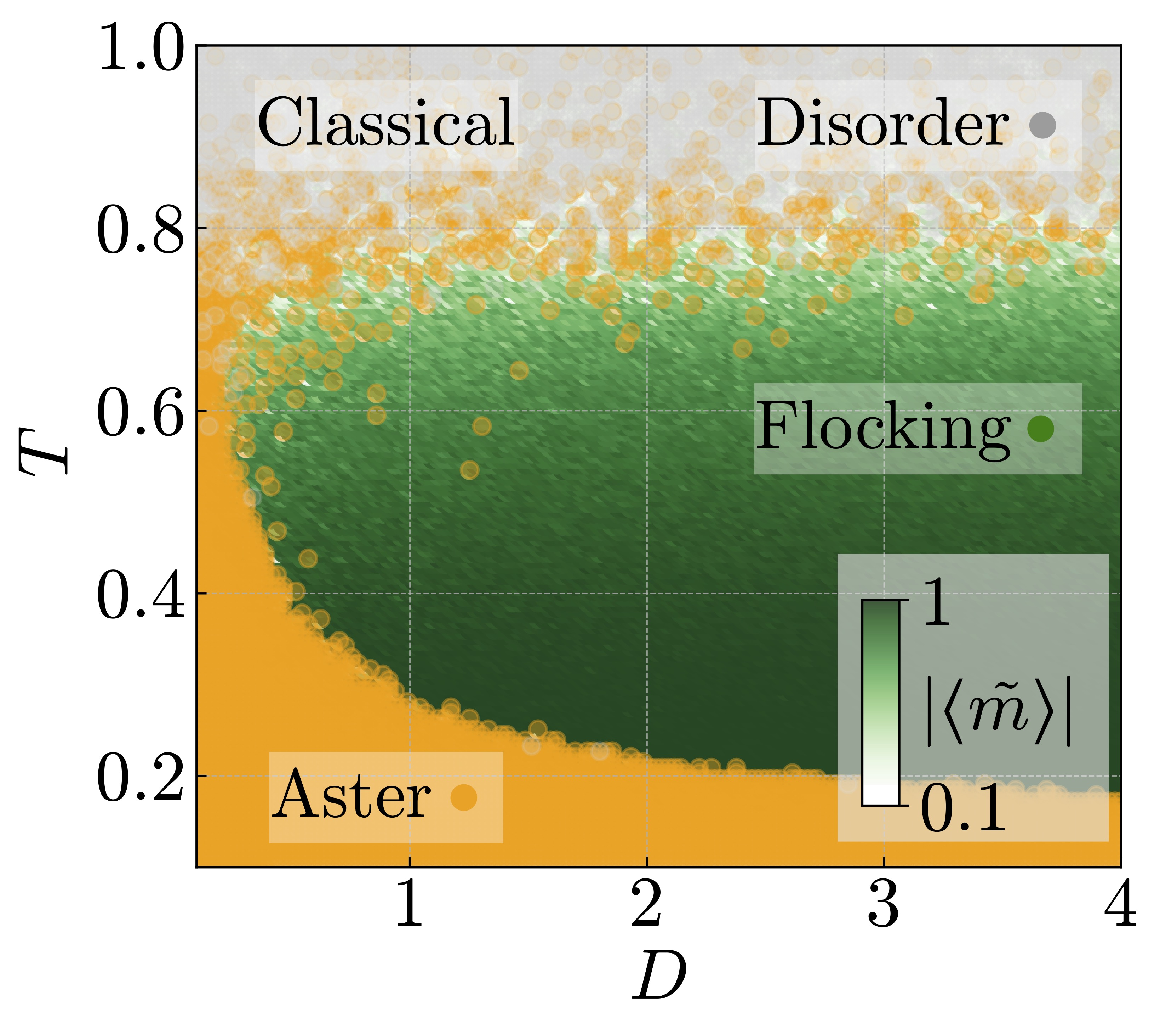}
    \label{mc:a}
  \end{subfigure}\hfill
    \begin{subfigure}[t]{0.24\textwidth}
    \centering
    \caption{\raggedright}
    \includegraphics[width=\textwidth]{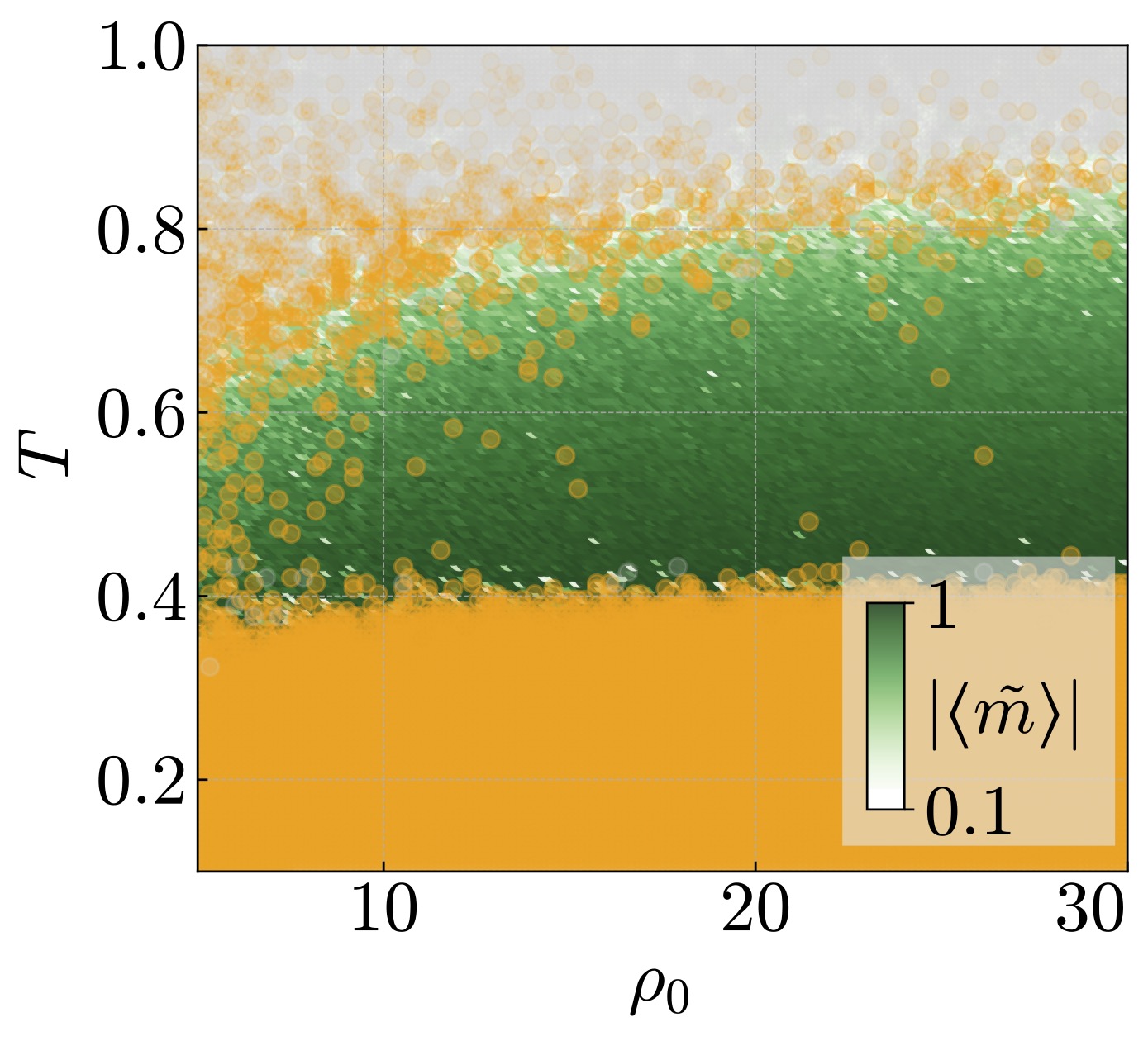}
    \label{mc:b}
  \end{subfigure}

  \vspace{-1.5em}
  \caption{\label{mc:ges}\justifying 
  Phase diagram of the classical active Ising model, \cref{master-classical}, (a) in the $T-D$ plane (for $\varepsilon = 0.7$, $\gamma = 1$,  $\rho_0 = 10$ and  $L=100$) and (b) in the $T-\rho_0$ plane for  $D=1$ (other parameters as in (a)). Yellow dots indicate the formation of asters (according to the critierion given the main text). The results are consistent with  \cite{classical_flocking}.}
  \label{mc_sim}
  \vspace{-1.5em}
\end{figure}

\subsection{Phase structure}
\label{eq:aim-class-phases}
To allow for direct comparison of our quantum results later, here we reproduce the phase diagram of the AIM that has previously been discussed in the literature \cite{classical_flocking}.  Instead of directly solving the classical master equation of the AIM, we solve the underlying Markov process via Monte Carlo methods by using the Gillespie algorithm \cite{monte_carlo,Vorberg2013,vorberg_nonequilibrium_2015}.  In the following, we set the amplitude of the spin flip rate to one, $\gamma=1$, that is, we use it as the natural unit for the rate $D$ and its inverse, $\gamma^{-1}$, as the natural unit for time.
The resulting phase diagrams in the $T-D$ and $T-\rho_0$ plane, respectively, where $\rho_0=N/L$ is the average particle density, are shown in \cref{mc_sim}, for an intermediate value $\varepsilon=0.7$ of the spin-dependent diffusion. We recover phase diagrams with identical transition temperatures to the ones that were previously found in Ref.~\cite{classical_flocking}.

When simulating the dynamics of the AIM,  for all parameters, we initialize the system in a disordered state by randomly assigning them a spin and a position on the lattice with uniform distribution (\cref{aim:a}).
The dynamics of the AIM can lead to the development of three different non-equilibrium phases (in 1D) \cite{classical_flocking}. 
These are, first, the disordered phase
with zero mean magnetization \emph{per particle} $\langle \tilde m \rangle = \sum_l m_l /N =0$ 
in the thermodynamic limit, and no pronounced local density variations. As expected, the disordered phase is stable at high temperatures, where the spin-spin interactions, which drive the collective behaviour, become irrelevant. 

Second, when the strength of mutual alignment increases, i.e., $T$ decreases, spin alignment becomes relevant. In presence of biased diffusion, that is, for sufficiently large $D$ and $\varepsilon$, the particles then gather
 in large flocks consisting of particles of the same species collectively moving in the same direction (\cref{aim:c}). This is indicated by a non-zero mean magnetization $\langle m \rangle=\sum_l m_l /L \neq 0$ of the system, with spontaneously chosen sign. Thus, the mean magnetization $\langle  m\rangle$ can be used as an order parameter. Starting the simulation from a disordered phase, $\langle \tilde m\rangle$ rises  quickly and reaches a plateau once a main flock has established itself. In the following, we use the threshold $|\langle\tilde{m}\rangle|>0.1$ to identify the flocking phase (of course, the precise value of 0.1 is somewhat arbitrary).
As seen from \cref{mc_sim}, the flocking phase is stable for a broad range of temperatures. 
The flocking phase becomes stabilized (in the sense of an increasing transition temperature from the high-temperature phase) when the mean density increases.

The third phase occuring in the 1D AIM is the so-called aster phase \cite{classical_flocking}.
Instead of continued propagation of particles through the AIM, an indicator of this phase is that two neighbouring sites become occupied by a large amount of same-species particles with opposite spin, so that spin-$\uparrow$ particles sit on the left of spin-$\downarrow$ particles, so that they hinder each others motion (\cref{aim:d}). In this phase, the mean magnetization vanishes, $\langle m \rangle  =0$, but there exists a large local  magnetisation with opposite sign on the two adjacent sites. 
To detect this phase numerically,
Ref.~\cite{classical_flocking} has proposed the following method:
One searches for two adjacent sites, $l$ and $l+1$, of opposite magnetization. 
If this is the case and the densities on both sites, $\rho_l$ and $\rho_{l+1}$, are, respectively, two times larger than the system's average density $\rho_0=N/L$, then we consider the system to be in an aster phase.
As seen from \cref{mc_sim}, the aster phase is stable at low temperatures and not too large values of $D$. 
To understand this behaviour, we can view asters as frustrated structures where two populations of right- and left-moving particles on adjacent lattice sites ``block" each other. 
Namely, particles constantly hop from one aster-site to the oppositely polarized opposing aster site, where they correspond to the minority and are, thus, rapidly spin flipped. 
At low temperatures and small $D$, these configurations are stable since periodic hopping and accompanying spin-flip processes (induced by strong alignment interactions) between the adjacent sites are much more likely than processes where particles hop to other sites. Therefore, individual asters have extremely long lifetimes (see \cite{classical_flocking} for a detailed analysis). Increasing $D$ (at fixed low $T$) leads to an increasing mobility of the particles. This disfavours the frustrated aster configurations and finally leads to their dissolution. Instead, the particles form large flocks.

\section{Quantum Master Equation for Active Bosons}\label{quantum_master_eq}
To describe active quantum matter out of equilibrium, we will use  
the formalism of open quantum systems. 
The stochastic dynamics of classical Markov processes is governed by classical master equations. The generalization of this master equation to time-homogeneous quantum Markov processes are master equations of Gorini–Kossakowski–Sudarshan–Lindblad (GKSL or simply Lindblad) form \cite{Breuer2002a,Weiss2012}
 \begin{align}
    &\timederiv \rho(t) 
    = - \frac{i}{\hbar} [H,\rho(t)] + \sum_{i}\mathcal{D}[L_i]\superopbr{\rho (t)}
    \equiv \mathcal{L} \superopbr{\rho(t)}.
    \label{eq:Lind-gen}
\end{align}
Here, $\rho(t)$ is the density matrix of the system, $H$ is the Hamiltonian governing the coherent quantum dynamics, and $\mathcal{D}[L_i]\superopbr{\rho (t)}$ are dissipators of the form
 \begin{align}
    \mathcal{D}[L_i]\superopbr{\rho} =  L_i \rho  L_i^\dagger - \frac{1}{2} \Bigl\{ L_i^\dagger L_i , \rho \Bigl\},
\end{align}
where $L_i$ are the so-called Lindblad jump operators describing  incoherent (bath-induced) processes in the system.

Since in our quantum active Ising model multiple terms of different physical origin enter, we write the total master equation as
 \begin{align}
    \mathcal{L}&= \mathcal{L}_\mathrm{coh} +  \mathcal{L}_\mathrm{F}+\mathcal{L}_\mathrm{H},
    \label{eq:quant-act-Liouv}
\end{align}
with $\mathcal{L}_\mathrm{coh}\superopbr{\rho(t)} =-i[H,\rho(t)]$ describing the coherent evolution, and with $\mathcal{L}_\mathrm{F}$ and $\mathcal{L}_\mathrm{H}$ describing incoherent spin-flip and hopping processes, respectively, as described in detail below. 

In Ref.~\@\cite{quantum_flocking}, the authors propose jump operators $L_i$  that generalize the classical AIM to hard-core bosons, and later include a Hamiltonian describing the Zeemann coupling to a magnetic field pointing in $x$-direction, that is transverse to the spin-$z$-axis along which the ferromagnetic Ising coupling is assumed. Accordingly, this field induces quantum fluctuations that counteract the dissipative ferromagnetic spin alignment.   Due to the investigation of hard-core bosons, the flipping rate in Ref.~\cite{quantum_flocking} was chosen slightly differently from the classical rate in \cref{rate-flip}, with the rate depending on the magnetization on neighboring sites.
Here, however, since we have ideal bosons, we can choose the rate in close analogy with \cref{rate-flip}.
This results in two types of dissipators, depending on whether a particle flips from $s=\uparrow$ to $s=\downarrow$ or vice versa, i.e.
\begin{align}
    \mathcal{L}_\mathrm{F} &= \mathcal{L}_{1}+ \mathcal{L}_{2},\\
    \mathcal{L}_{1/2}\superopbr{\rho}&= \gamma \sum_{l=1}^L \mathcal{D}\Bigl[a_{l \uparrow/\downarrow }^\dagger a_{l \downarrow/\uparrow} e^{\pm \frac{\beta^*}{2} (n_{l \uparrow}-n_{l \downarrow})}\Bigl]\superopbr{\rho}.
    \label{eq:flip-incoh}
\end{align}
Here enter the occupation number operators $n_{ls} = a_{ls}^\dagger a_{ls}$.
Note that, since later we want to perform a mean-field expansion, we require a polynomial expansion in the number operator. 
Hence in our model with inverse temperature $\beta^*$, we omit the inclusion of the total particle density $n_{l\uparrow} + n_{l\downarrow}$  in the denominator of the exponent in the rate \cref{rate-flip}. 
We expect that this will lead to an effective shift of the temperature of the phase transition to the Flocking phase, when compared to the classical active Ising model. In order to stabilize the numerics (by avoiding exponentially large rates), and  to make our results comparable to the classical active Ising model, we will later reintroduce the missing factor in an ad-hoc fashion by replacing 
\begin{equation}
\beta^* \rightarrow \frac{\beta}{\langle n_{l\uparrow} \rangle + \langle n_{l\downarrow}\rangle},
\label{eq:beta-star}
\end{equation} which forces the exponent in the rate in \cref{eq:flip-incoh} to be on the order of $\beta$ as used in similar models \cite{classical_flocking, Cates2015, Bandyopadhyay_2024}.
We note in this context that, at least for the classical AIM, first, there is some freedom in the precise definition of the spin-flip rate as long as it favors alignment \cite{scandolo_active_2023}, and second, also in the classical model we observe that dividing by the mean density is required to make the numerics stable.


Next, we define dissipators for the biased diffusion: 
Here, the probability of a particle hopping to an adjacent site depends on a basic incoherent hopping rate $D$ as well as the activity $\varepsilon$ of the medium, resulting in two operators for each direction of hopping
\begin{align}
\mathcal{L}_\mathrm{H} &= \mathcal{L}_\mathrm{H,r}+\mathcal{L}_\mathrm{H,l},
\intertext{with}
\mathcal{L}_\mathrm{H,r} &= \mathcal{L}_3 + \mathcal{L}_6, \\
    \mathcal{L}_{3/6}\superopbr{\rho}&=D (1 \pm \varepsilon) \sum_{l=1}^L \mathcal{D}\Bigl[a^\dagger_{l+1 \uparrow/\downarrow} a_{l \uparrow/\downarrow}\Bigl]\superopbr{\rho},
\intertext{and}
\mathcal{L}_\mathrm{H,l} &= \mathcal{L}_4 + \mathcal{L}_5, \\
    \mathcal{L}_{4/5}\superopbr{\rho}&=D (1 \pm \varepsilon) \sum_{l=1}^L \mathcal{D}\Bigl[a^\dagger_{l-1 \downarrow/\uparrow} a_{l \downarrow/\uparrow}\Bigl]\superopbr{\rho}.
\end{align}
In order to study the influence of quantum fluctuations on the active matter system, we also introduce a transverse  external magnetic field  via the Hamiltonian
\begin{align}
    H&=\hbar \omega \sum_{l=1}^L \left(a^\dagger_{l\uparrow} a_{l\downarrow} + a^\dagger_{l\downarrow} a_{l\uparrow} \right)
\end{align}
which rotates the spins coherently around the $x$ axis with angular frequency $\omega$.  
Here $\omega$ determines the strength of the field. 



\subsection{ Kinetic equations: General remarks}

Since the dimension of the Hilbert space of the bosonic many-body system is growing extremely fast when approaching the thermodynamic limit $L,N\to\infty$ (while keeping $\rho_0=N/L$ fixed), solving the full dynamics of the Liouvillian in \cref{eq:quant-act-Liouv} can only be done for relatively small chains and low particle numbers. To obtain solutions for large systems, in the following sections, we will 
derive a closed set of non-linear kinetic equation of motion for the single particle density matrix $\langle a_{ls}^\dag a_{l’s’}\rangle$ starting from 
\begin{align}
\timederiv \langle a^\dag_{ls}a_{l's'}\rangle = \text{tr}(a^\dag_{ls}a_{l's'}\timederiv  \rho(t))
\label{eq:eom-spdmat}
\end{align}
and by employing a mean field approximation.
In particular, the diagonal elements of the single-particle density matrix correspond to the mean-occupations of spin-$s$ particles on site $l$,  
\begin{align}
     \langle n_{ls} \rangle 
    = \langle a^\dag_{ls}a_{ls}\rangle.
    \label{trace}
\end{align}

When plugging the master equation defined by \cref{eq:Lind-gen,eq:quant-act-Liouv}
into \cref{eq:eom-spdmat}, one finds that the evolution of the single-particle density operator depends on two- and more-particle correlations (given by expectation values of products of four or more bosonic annihilation and creation operators). 
This corresponds to a BBGKY-type hierarchy of equations of motion, as it indicates the presence of interactions. In order to get  a closed set of equations of motion, we can decompose the two-particle correlations on the right-hand-side of Eq.~(\ref{eq:eom-spdmat}) into products of single-particle occupations, assuming Wick's theorem to hold. 
This assumption is equivalent to the ansatz that the system's state is Gaussian, $\rho \propto \exp(-\sum_{l,l’}\sum_{s,s’} \eta_{ls,l’s’}a^\dag_{ls}a_{l’s’})$ with coefficients $\eta_{ls,l’s’}^*=\eta_{l’s’,ls}$. For bosonic Gaussian states, it predicts for two-particle correlators that 
\begin{align}
    \langle a b c d\rangle = \langle ab \rangle \langle cd\rangle +\langle ac \rangle \langle bd \rangle + \langle ad \rangle \langle bc\rangle, \label{eq:wick-d}
\end{align}
where $a, b, c, d$ are arbitrary bosonic annihilation or creation operators. As the dissipative terms of the Liouvillian are quartic in the bosonic annihilation operators (reflecting the fact that they are the result of interactions between system particles and environment), this is not exact, but corresponds to a mean-field approximation \cite{vorberg_nonequilibrium_2015,schnell_number_2018,lange_pumping_2017}.
In principle, this leads to $(2 L)^2$ nonlinear kinetic equations which allow us to solve relatively large systems numerically.

Note that in case of vanishing external magnetic field, we can additionally neglect coherences between different spin states and different 
sites, i.e., terms of the form $\langle a^\dagger_{ls}a_{l's'}\rangle$ with $(l,s) \neq (l',s')$.
Namely, if such coherences are not built up by the Hamiltonian,
they will decay in time and vanish  in the steady state. In the theory of open quantum systems, this is known as dynamical decoupling and can be seen most easily by looking at a simple model of Lindblad jump operators that only induce jumps between eigenstates $\ket{i}$,
\begin{align}
    \timederiv \rho(t) = \sum_{ij} R_{ij} \mathcal{D}[\ket{i}\bra{j}]\superopbr{\rho(t)},
\end{align}
with corresponding jump rate $R_{ij}$. This leads to equations of motion of the matrix elements of the density matrix reading \cite{Breuer2002a}
\begin{align}
    \timederiv \rho_{ii'}(t) = \left\lbrace \begin{array}{cc}
    \sum_j \left[R_{ij} \rho_{jj}(t)-R_{ji} \rho_{ii}(t)\right], & i = i',\\
    -\frac{1}{2}\sum_j \left(R_{ji} + R_{ji'}\right) \rho_{ii'}(t), & i \neq i'.
    \end{array}\right.
    \label{eq:pauli-and-coh}
\end{align}
While the coherences 
(the off-diagonal elements of the density matrix)
decay in time, the populations (the diagonal elements) are governed by a Pauli master equation.
For our model without external field, $\omega=0$, we can associate the states $i$ and $j$ with the Fock states $\ket{\mathbf{n}} = \ket{n_{1\uparrow}, n_{1\downarrow}, \dots}$. This means that at long times the density matrix will be diagonal with respect to these Fock states.
Hence, in absence of a magnetic field, we can neglect coherences of the form $\langle a^\dagger_{ls}a_{l's'}\rangle$ with $(l,s) \neq (l',s')$. 

Before describing the detailed rates for our bosonic active quantum chain, let us already point out one important aspect. Assuming the rates for single particles jumping from state a state $q$ to a state $k$ to be given by $R_{kq}$, we obtain the corresponding many-boson problem by substituting the corresponding jump operators according to the rule $|k\rangle\langle q|\to a^\dag_k a_q$, giving rise to the Liouvillian $\mathcal{L}=\sum_{kq}R_{kq}\mathcal{D}[a_k^\dag a_q]$. This leads to the equations of motions for the mean-occupations (see, e.g., Ref.~\cite{vorberg_nonequilibrium_2015} and detailed derivation given below) 
\begin{align}
\begin{split}
\timederiv \langle n_{k} \rangle 
    & = \text{tr}(n_{k}\timederiv  \rho(t))
  \\
    &= \sum_{q\ne k} \big[R_{kq}\langle(1+n_k)n_q\rangle - R_{qk}\langle (1+n_q)n_k\rangle\big],
\end{split}
    \label{eq:NoCoh}
\end{align}
where in our system $k=(l,s)$ and $q=(l',s')$. This equation directly corresponds to the classical rate equation in \cref{eq:pauli-and-coh}, with one important exception: Namely the rate for a boson jumping from state $k$  to state $q$ is given by
\begin{equation}
R_{qk}\langle (1+n_q)n_k\rangle\approx R_{qk}\langle 1+n_q\rangle\langle n_k\rangle
\end{equation}
(where the factorization results from the mean-field approximation). We can see that it does not only depend on the number of particles $\langle n_k\rangle$ occupying the initial state, but also, via the factor $\langle 1+n_q\rangle$, on the number of particles in the final state $q$. This is a quantum statistical effect known as bosonic enhancement. It reflects the tendency of indistinguishable bosons to occupy the same state and gives rise, e.g., also to stimulated emission. It is absent in the classical model, which describes distinguishable particles. Thus, already without considering quantum fluctuations, as they are induced by the transverse magnetic field, the generalization to a quantum system gives rise to a new effect that we expect to enhance the collective processes and stabilize collective behaviour of the system.

Note that in the absence of coherences, we can efficiently simulate the dynamics of \cref{eq:NoCoh} also without mean-field approximation by generating random walks between the bosonic Fock states using the Gillespie algorithm \cite{vorberg_nonequilibrium_2015}.
In the following, we will, however, derive and solve kinetic equations. 

\subsection{Detailed derivation of the kinetic equations for the dissipators}\label{kian_master_2}

 To derive the kinetic equations for the mean occupations (i.e.\ in the absence of the magnetic field), we start by defining
\begin{align}
\timederiv \langle n_{k\uparrow} \rangle_i =  \text{tr}\left(n_{k\uparrow} \mathcal{L}_i\superopbr{\rho(t)}\right),
\end{align}
as the part of the change of the occupations due to the Liouvillian $\mathcal L_i$. 

For $\omega=0$, the time evolution generated by the hopping processes is given by 
\begin{align}
\timederiv \langle n_{k\uparrow} \rangle_\mathrm{H} = \sum_{i=3}^6 \timederiv \langle n_{k\uparrow} \rangle_i.
\end{align}
To evaluate the right-hand side, let us first note that for a general observable $O$  
one finds that 
\begin{align}
\begin{split}
\mathrm{tr}\left\{ O\mathcal{D}[L]\superopbr{\rho}\right\}
&= \mathrm{tr}\left( O L\rho L^\dagger- \frac{1}{2} O \lbrace L^\dagger L,\rho  \rbrace\right)\\
& = \frac{1}{2} \langle L^\dagger [O, L] \rangle + \frac{1}{2} \langle [ L^\dagger, O] L \rangle.
\end{split}
\end{align}
Here we have used cyclic permutations under the trace to identify expectation values.
With this, we calculate 
\begin{align}
\begin{split}
\timederiv \langle n_{k\uparrow} \rangle_3  = 
D (1& + \varepsilon)  \sum_{l=1}^L\left(\frac{1}{2} \left\langle a_{l \uparrow}^\dagger a_{l+1 \uparrow} \left[n_{k\uparrow},  a^\dagger_{l+1 \uparrow} a_{l \uparrow}\right]\right\rangle\right.
\\
&+\left.\frac{1}{2} \left\langle \left[a_{l \uparrow}^\dagger a_{l+1 \uparrow} ,n_{k\uparrow} \right] a^\dagger_{l+1 \uparrow} a_{l \uparrow}\right\rangle \right).
\end{split}
\end{align}
Employing
\begin{align}
 [n_{k\uparrow},  a^\dagger_{l+1 \uparrow} a_{l \uparrow}] = \delta_{k,l+1} a^\dagger_{k\uparrow}a_{l \uparrow} -\delta_{kl} a^\dagger_{l+1 \uparrow} a_{k \uparrow},
\end{align}
using that the second term is the hermitian conjugate of the first one, and reordering the bosonic annihilation and creation operators, one finds
\begin{align}
\begin{split}
\timederiv &\langle n_{k\uparrow} \rangle_3 = \\
&D (1 + \varepsilon)  \left[ \left\langle (1+n_{k \uparrow}) n_{k-1 \uparrow}\right\rangle
-\left\langle (1+n_{k+1\uparrow}) n_{k \uparrow} \right\rangle \right].
\end{split}
\label{eq:disshop-right}
\end{align}
This expression is rather intuitive since, e.g., the first term corresponds to a jump from site $k-1$ to $k$ where the rate depends on the occupation $n_{k-1\uparrow}$  of the original site (as we would expect in a classical system) as well as the occuaption of the target site via the bosonic enhancement factor $1+n_{k \uparrow}$, which stems from the indistinguishability of the bosons on the target site.
Analogously, we can calculate $\timederiv \langle n_{k\uparrow} \rangle_5 $.
Since the jump operators in $\mathcal L_4,\mathcal L_6$ act on the down spin component only, we have, moreover, $\timederiv \langle n_{k\uparrow} \rangle_{4,6} = 0$.
As discussed earlier, we observe that the change of the mean occupation depends on two-particle correlations.
By applying a Wick decomposition [\cref{eq:wick-d}] and using that the particle number is conserved during the evolution, we find the part of the kinetic equation stemming from the dissipative hopping  to be given by
\begin{align}
    \begin{split}
    \timederiv \langle & n_{k\uparrow} \rangle_\mathrm{H} = \\
    D &(1 +  \varepsilon) \Big( 
     \langle 1+ n_{k\uparrow} \rangle  \langle n_{k-1\uparrow} \rangle
      - \langle 1+ n_{k+1\uparrow} \rangle \langle n_{k\uparrow} \rangle \\ 
    &+ \langle a^\dagger_{k-1\uparrow} a_{k\uparrow} \rangle \langle a^\dagger_{k\uparrow} a_{k-1\uparrow} \rangle 
    - \langle a^\dagger_{k+1\uparrow} a_{k\uparrow} \rangle \langle a^\dagger_{k\uparrow} a_{k+1\uparrow} \rangle 
    \Big) \\
    + & D (1 - \varepsilon) \Big( 
    \langle 1+ n_{k\uparrow} \rangle\langle n_{k+1\uparrow} \rangle 
    - \langle 1+  n_{k-1\uparrow} \rangle  \langle n_{k\uparrow} \rangle \\
    &+ \langle a^\dagger_{k+1\uparrow} a_{k\uparrow} \rangle \langle a^\dagger_{k\uparrow} a_{k+1\uparrow} \rangle 
    - \langle a^\dagger_{k-1\uparrow} a_{k\uparrow} \rangle \langle a^\dagger_{k\uparrow} a_{k-1\uparrow} \rangle 
    \Big).
    \end{split}
    \label{coherent_hop}
\end{align}
Note that without magnetic field, $\omega =0$, the terms of the form $\langle a^\dagger_{k-1\uparrow} a_{k\uparrow} \rangle$ which describe coherences between neighboring sites vanish (due to the dynamical decoupling discussed above).
In that case, we have
\begin{align}
    \begin{split}
    \timederiv \langle & n_{k\uparrow} \rangle_\mathrm{H} = \\
    D &(1 +  \varepsilon) \Big( 
     \langle 1+ n_{k\uparrow} \rangle  \langle n_{k-1\uparrow} \rangle
      - \langle 1+ n_{k+1\uparrow} \rangle \langle n_{k\uparrow} \rangle 
    \Big) \\
    + & D (1 - \varepsilon) \Big( 
    \langle 1+ n_{k\uparrow} \rangle\langle n_{k+1\uparrow} \rangle 
    - \langle 1+  n_{k-1\uparrow} \rangle  \langle n_{k\uparrow} \rangle 
    \Big).
    \end{split}
\end{align}
Analogously, we find 
\begin{align}
    \begin{split}
    \timederiv \langle & n_{k\downarrow} \rangle_\mathrm{H} = \\
    D &(1 -  \varepsilon) \Big( 
     \langle 1+ n_{k\downarrow} \rangle  \langle n_{k-1\downarrow} \rangle
      - \langle 1+ n_{k+1\downarrow} \rangle \langle n_{k\downarrow} \rangle 
    \Big) \\
    + & D (1 + \varepsilon) \Big( 
    \langle 1+ n_{k\downarrow} \rangle\langle n_{k+1\downarrow} \rangle 
    - \langle 1+  n_{k-1\downarrow} \rangle  \langle n_{k\downarrow} \rangle 
    \Big).
    \end{split}
\end{align}
for spin-$\downarrow$-particles. 

In presence of the magnetic field, however, on-site coherences  $\langle a^\dagger_{k\uparrow} a_{k\downarrow} \rangle$ are built up. 
Nevertheless, we expect that the coherences as $\langle a^\dagger_{k-1\uparrow} a_{k\uparrow} \rangle$  between \emph{different} sites will  still vanish in the steady state.  Namely, the hopping terms are dissipative, so that they do not build up coherences between different sites. Moreover, they include a projection on the spin state, which hinders the propagation of on-site coherence between different spin states to particles on different sites. 

In the next step, we study the change of the mean occupations resulting from the flipping terms,

\begin{align}
\timederiv \langle n_{k \uparrow}\rangle_\mathrm{F} = \timederiv \langle n_{k \uparrow}\rangle_1 +\timederiv \langle n_{k \uparrow}\rangle_2.
\end{align}
Similar to above, we find 
\begin{align}
\begin{split}
\timederiv \langle &n_{k\uparrow} \rangle_1  =\\ 
\gamma &  \sum_{l=1}^L\left(\frac{1}{2}\left\langle e^{ \frac{\beta^*}{2}  (n_{l \uparrow}-n_{l \downarrow})} a_{l \downarrow}^\dagger a_{l \uparrow} \left[n_{k\uparrow},  a^\dagger_{l \uparrow} a_{l \downarrow} e^{ \frac{\beta^*}{2}  (n_{l \uparrow}-n_{l \downarrow})}\right]\right\rangle\right.
\\
&+\left.\frac{1}{2} \left\langle \left[e^{ \frac{\beta^*}{2} (n_{l \uparrow}-n_{l \downarrow})} a_{l \downarrow}^\dagger a_{l \uparrow}, n_{k\uparrow}  \right]a^\dagger_{l \uparrow} a_{l \downarrow} e^{ \frac{\beta^*}{2}  (n_{l \uparrow}-n_{l \downarrow})}\right\rangle \right),
\end{split}
\end{align}
for the dissipative  flip from down to up.
Here, evaluating the commutator gives
\begin{align}
\left[n_{k\uparrow},  a^\dagger_{l \uparrow} a_{l \downarrow} e^{ \frac{\beta^*}{2}  (n_{l \uparrow}-n_{l \downarrow})}\right] &= \left[n_{k\uparrow},  a^\dagger_{l \uparrow}\right] a_{l \downarrow} e^{ \frac{\beta^*}{2}  (n_{l \uparrow}-n_{l \downarrow})}\\
&=\delta_{kl}a^\dagger_{k \uparrow} a_{k \downarrow} e^{ \frac{\beta^*}{2}  (n_{k \uparrow}-n_{k \downarrow})}.
\end{align}
After reordering the bosonic operators, this finally yields
\begin{align}
\timederiv \langle n_{k\uparrow} \rangle_1  = 
\gamma &  \langle(1+n_{k \uparrow}) e^{ \beta^* n_{k \uparrow}}  n_{k \downarrow} e^{ -\beta^* n_{k \downarrow}}\rangle.
\label{eq:eom-flip1}
\end{align}
Again, we observe that the rate of change of occupation of the up-spin species on site $k$ depends on the occupation of the down spin on this site and additionally the occupation of the up spin according to bosonic enhancement. Moreover, there are the Boltzmann factors which increases the rate of the dissipative flips into the spin up state if its population is higher than in the down state, $n_{k \uparrow} > n_{k \downarrow}$. Hence the Boltzmann factors give rise to spin alignment.

\begin{figure}[t]
\includegraphics[width=0.5\textwidth]{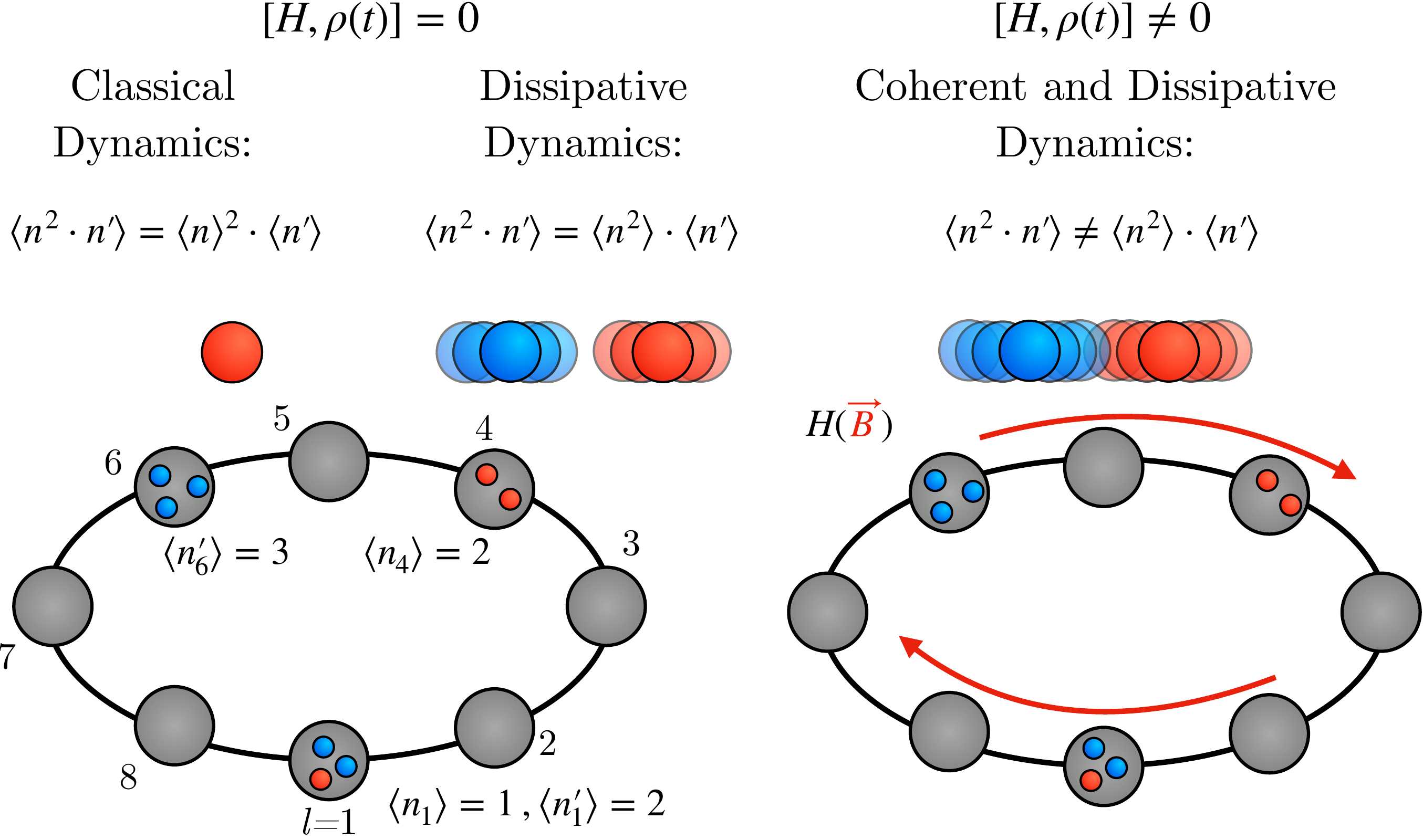}
  \vspace{-0.5em}
\caption{\label{dynamics}\justifying Different setups of the AIM under investigation and the corresponding master equations.
  In the classical limit, all quantum effects are neglected when evaluating the expectation values. 
  When focusing on dissipative quantum effects, the coherences resulting from a separation of $\langle n \cdot n'\rangle$ are set to zero.
  Lastly, to simulate an external influence, in our case a magnetic field, the system is exposed to an external Hamiltonian $H(\vec{B})\neq 0$. The resulting dynamics include quantum coherences.}
  \label{fig:coherences-sketch}
\end{figure}

In complete analogy, for the reverse flip, we have
\begin{align}
\timederiv \langle n_{k\uparrow} \rangle_2  = -
\gamma &  \langle(1+n_{k \downarrow}) e^{ \beta^* n_{k \downarrow}}   n_{k \uparrow} e^{ -\beta^* n_{k \uparrow}}\rangle.
\label{eq:eom-flip2}
\end{align}
In absence of an external magnetic field, $\omega=0$, there are no coherences between the up- and downspin components (cf.~\cref{fig:coherences-sketch}). Hence for $\omega=0$, we have
\begin{align}
\begin{split}
\timederiv \langle n_{k\uparrow} \rangle_\mathrm{F}  = 
& \gamma   \langle(1+n_{k \uparrow}) e^{ \beta^* n_{k \uparrow}}  \rangle \langle n_{k \downarrow} e^{ -\beta^* n_{k \downarrow}}\rangle\\
 & -
\gamma  \langle(1+n_{k \downarrow}) e^{ \beta^* n_{k \downarrow}}  \rangle \langle n_{k \uparrow} e^{ -\beta^* n_{k \uparrow}}\rangle.
\end{split}
\label{eq:flip-nocoh}
\end{align}
Note that for Gaussian states, the higher moments of the occupation numbers can be reduced to functions of the mean occupation.
As we discuss in Appendix~\ref{appendix_a}, expanding the exponential in a polynomial series and applying Wick's theorem on each order of the particle-particle correlations, we obtain
\begin{align}
    \big\langle e^{\alpha n} \big\rangle &= 1+\sum_{k=1}^{\infty}\frac{\alpha^k}{k!} \left(\sum_{q=1}^{k} \begin{Bmatrix}k \\ q \end{Bmatrix} q!\langle n \rangle^q \right) \label{en-main}\\
    \big\langle n e^{\alpha n} \big\rangle &= \langle n \rangle +\sum_{k=1}^{\infty}\frac{\alpha^k}{k!} \left(\sum_{q=1}^{k+1} \begin{Bmatrix}k+1 \\ q\end{Bmatrix} q!\langle n \rangle^q \right) \label{nen-main}
\end{align}
where $\begin{Bmatrix}
    k,q
\end{Bmatrix}^T$ are the Stirling numbers of the second kind. In this way, the expectation values are reduced to mean occupations and we truncate the series at a high order $k$ close to convergence. 
This allows us to investigate the dynamics depending on the expansion order and verify that often even a first order expansion properly resolves the phases of the system.\\
\\

In case of a finite magnetic field $\omega \neq 0$ (cf.~\cref{fig:coherences-sketch}) an analogous series up to arbitrary order of the exponential is not at hand. Hence, we will focus on the regime of high temperatures $\beta^* \langle n\rangle \ll 1$, where  we expand the exponentials in \cref{eq:eom-flip1} and \cref{eq:eom-flip2} up to first order in inverse temperature and then perform a Wick decomposition to find
\begin{align}
\begin{split}
    &\timederiv \langle n_{k\uparrow} \rangle_\mathrm{F} = \gamma 
    \Big[
      (1-\beta^*)(\langle n_{k\downarrow} \rangle - \langle n_{k\uparrow} \rangle) \\
    &-2\beta^* \left( \langle n_{k\downarrow} \rangle^2 - \langle n_{k\uparrow} \rangle^2 + 4 \langle a^\dagger_{k\downarrow} a_{k\uparrow} \rangle \langle a^\dagger_{k\uparrow} a_{k\downarrow} \rangle ( \langle n_{k\downarrow} \rangle - \langle n_{k\uparrow} \rangle)
    \right) \\
    &+4 \beta^*  \langle n_{k\downarrow} \rangle \langle n_{k\uparrow} \rangle \left( \langle n_{k\uparrow} \rangle-\langle n_{k\downarrow} \rangle \right) \Big].
    \label{coherent_flip}
\end{split}
\end{align}

All in all, in the absence of an external transverse magnetic field, $\omega=0$, we find the closed set of kinetic equations for the mean occupations 
\begin{align}
    \begin{split}
    \timederiv \langle & n_{k\uparrow} \rangle = \\
    D &(1 +  \varepsilon) \Big( 
     \langle 1+ n_{k\uparrow} \rangle  \langle n_{k-1\uparrow} \rangle
      - \langle 1+ n_{k+1\uparrow} \rangle \langle n_{k\uparrow} \rangle 
    \Big) \\
    + & D (1 - \varepsilon) \Big( 
    \langle 1+ n_{k\uparrow} \rangle\langle n_{k+1\uparrow} \rangle 
    - \langle 1+  n_{k-1\uparrow} \rangle  \langle n_{k\uparrow} \rangle 
    \Big)\\
    +& \gamma   \langle(1+n_{k \uparrow}) e^{ \beta^* n_{k \uparrow}}  \rangle \langle n_{k \downarrow} e^{ -\beta^* n_{k \downarrow}}\rangle\\
 -& 
\gamma  \langle(1+n_{k \downarrow}) e^{ \beta^* n_{k \downarrow}}  \rangle \langle n_{k \uparrow} e^{ -\beta^* n_{k \uparrow}}\rangle ,
    \end{split}
    \label{eq:n-mean-compl}
\end{align}
where the higher order correlations in the lower two lines have to be expanded using \cref{en-main} and \cref{nen-main}, and the corresponding equation for $\timederiv \langle  n_{k\downarrow} \rangle$ can be found by exchanging all spin up with spin down and vice versa in \cref{eq:n-mean-compl}. Before studying the case of non-zero transverse magnetic fields in \cref{Derivation} below, let us first discuss the relation to a classical system of distinguishable particles. 

\subsection{Relation to classical system}\label{kian_master_1}
Let us check, whether our   quantum model recovers the physics of the classical AIM in the case of classical (distinguishable) particles occupying the quantum AIM in absence of magnetic field, $\omega=0$.

For the incoherent hopping terms, e.g.~in \cref{eq:disshop-right}, this can be achieved simply by removing the bosonic enhancement factors
\begin{align}
\begin{split}
    \timederiv \langle n_{k\uparrow} \rangle_\mathrm{H} = & D (1 + \varepsilon) \Big( 
    \langle n_{k-1\uparrow} \rangle 
     - \langle n_{k\uparrow} \rangle  
    \Big) \\
    &+ D (1 - \varepsilon) \Big( 
    \langle n_{k+1\uparrow} \rangle 
    -  \langle n_{k\uparrow} \rangle 
    \Big).
\end{split}    
\label{incoherent_hop_class}
\end{align}
Similarly, by neglecting the bosonic enhancement factor in \cref{eq:eom-flip1} for the flipping part, we have
\begin{align}
\begin{split}
\timederiv \langle n_{k\uparrow} \rangle_\mathrm{F}  = 
 \gamma   \langle e^{ \beta^* (n_{k \uparrow}-n_{k \downarrow})}  n_{k \downarrow} \rangle -
\gamma  \langle e^{ \beta^* (n_{k \downarrow}-n_{k \uparrow})}   n_{k \uparrow} \rangle.
\end{split}
\label{eq:flip-nocoh-class}
\end{align}
Note that these two equations, alternatively, can also be found by writing the master equation of the corresponding AIM in the classical limit. 
This master equation is given in Eq.~(\ref{master-classical})
where we need to replace $\beta$ according to \cref{eq:beta-star} in the spin-flip rates.
By calculating the change of the mean occupation numbers from this master equation using
\begin{align}
    \timederiv \langle n_{k \uparrow} \rangle = \sum_{\mathbf{n}} n_{k \uparrow} \, \timederiv p_{\mathbf{n}},
\end{align}
we reproduce \cref{incoherent_hop_class} and \cref{eq:flip-nocoh-class}.

Note that the exponential factors on the right hand side of \cref{eq:flip-nocoh-class} require us to also have knowledge of the higher moments of the occupation numbers $n_{ks}$. 
For the \emph{indistinguishable bosons}, using Wick's theorem, we find for the two-particle correlations in absence of coherences
\begin{align}
 \langle n_{ks} n_{k' s'} \rangle \overset{\text{bosons}}{=} \left\lbrace \begin{array}{cc}
 2 \langle n_{ks}\rangle^2 + \langle n_{k s} \rangle  & \text{for } k = k', s = s',\\
  \langle n_{ks}\rangle \langle n_{k' s'} \rangle  & \text{else}.
\end{array} 
 \right.
\end{align}
To find the corresponding behavior for \emph{distinguishable classical particles}, we take the expression for distinguishable states $k \neq k'$ and also apply it to the case $k = k'$.
\begin{align}
 \langle n_{ks} n_{k' s'} \rangle \overset{\text{class. particles}}{=} 
  \langle n_{ks}\rangle \langle n_{k' s'} \rangle. 
\end{align}
Repeating this argument, we find that the higher moments decompose trivially  $\langle n_{ks}^i \rangle = 
  \langle n_{ks}\rangle^i $. 
In this way, we arrive at closed equations of motion for classical particles, reading
\begin{align}
\begin{split}
\timederiv \langle & n_{k\uparrow} \rangle =  D (1 + \varepsilon) \Big( 
    \langle n_{k-1\uparrow} \rangle 
     - \langle n_{k\uparrow} \rangle  
    \Big) \\
    &+ D (1 - \varepsilon) \Big( 
    \langle n_{k+1\uparrow} \rangle 
    -  \langle n_{k\uparrow} \rangle 
    \Big)\\
    &+ \gamma    e^{ \beta^* (\langle n_{k \uparrow}\rangle-\langle n_{k \downarrow}\rangle)} \langle n_{k \downarrow} \rangle -
\gamma   e^{ \beta^* (\langle n_{k \downarrow}\rangle-\langle n_{k \uparrow}\rangle)}  \langle n_{k \uparrow} \rangle.
\end{split}
\label{eq:kineq-classical}
\end{align}
These equations correspond to a mean-field approximation to the classical AIM in \cref{master-classical}. However, we note that the here presented mean-field approach in the classical limit is different in character compared to the mean-field theory of the classical AIM presented in Ref.~\cite{classical_flocking}. There, the  mean-field approximation enters only after taking the continuum limit of the kinetic equations (that have the same form as our Eqs.~\eqref{incoherent_hop_class}, \eqref{eq:flip-nocoh-class}), resulting in mean-field equations for the spatiotemporal density- and magnetization fields. Here we rather consider a site-resolved mean-field approach.
We will see below that the analysis of the classical AIM based on our kinetic lattice equations \eqref{eq:kineq-classical} allows to describe all three phases.



\subsection{Kinetic equations for non-zero transverse field}\label{Derivation}
\begin{figure*}[t]
  \centering
  \begin{subfigure}[t]{0.35\textwidth}
    \centering
    \caption{\raggedright}
    \includegraphics[width=\textwidth]{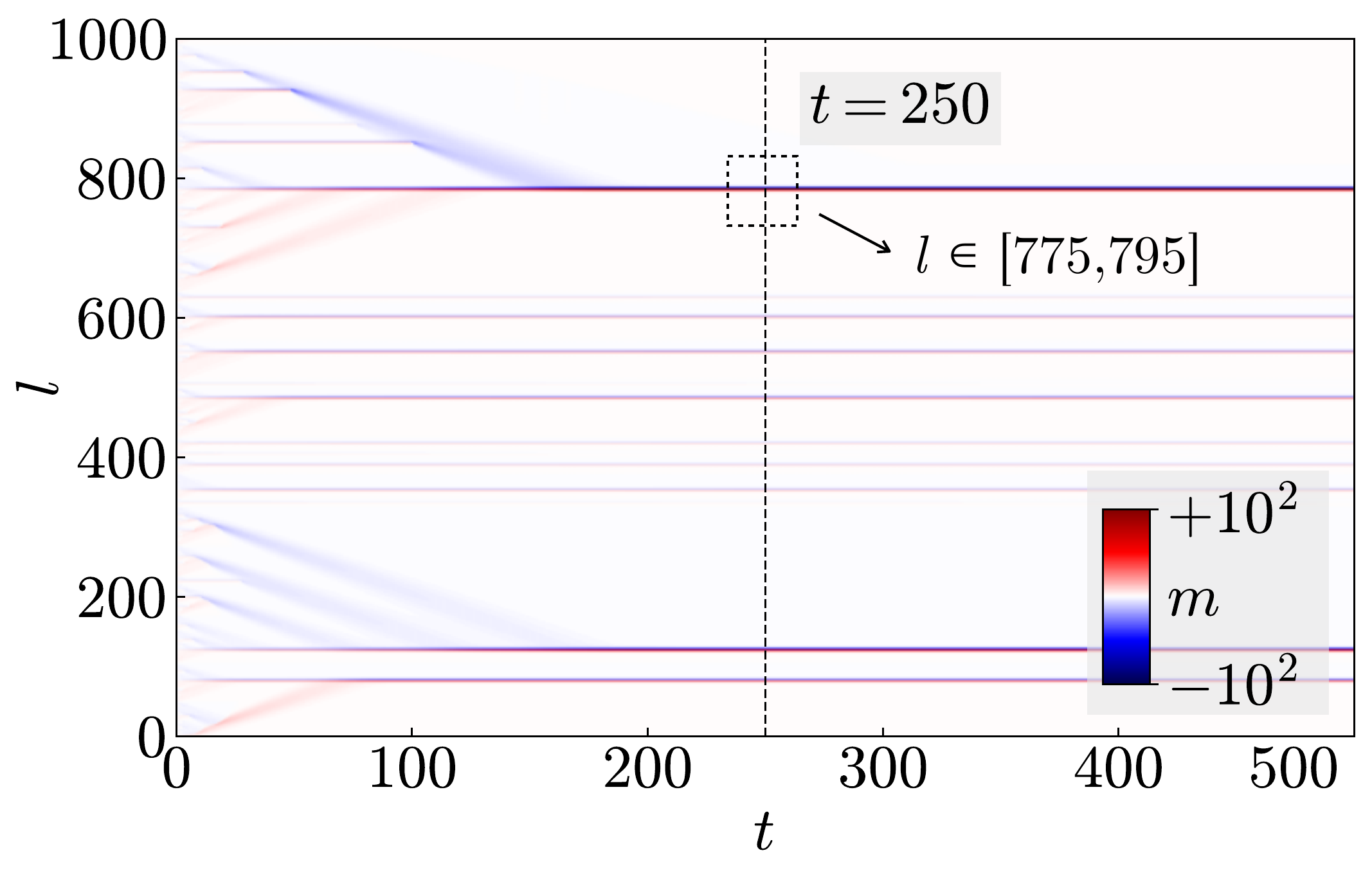}
    \label{cl:a}
  \end{subfigure} \hspace{0.3cm}
    \begin{subfigure}[t]{0.37\textwidth}
    \centering
    \caption{\raggedright}
    \includegraphics[width=\textwidth]{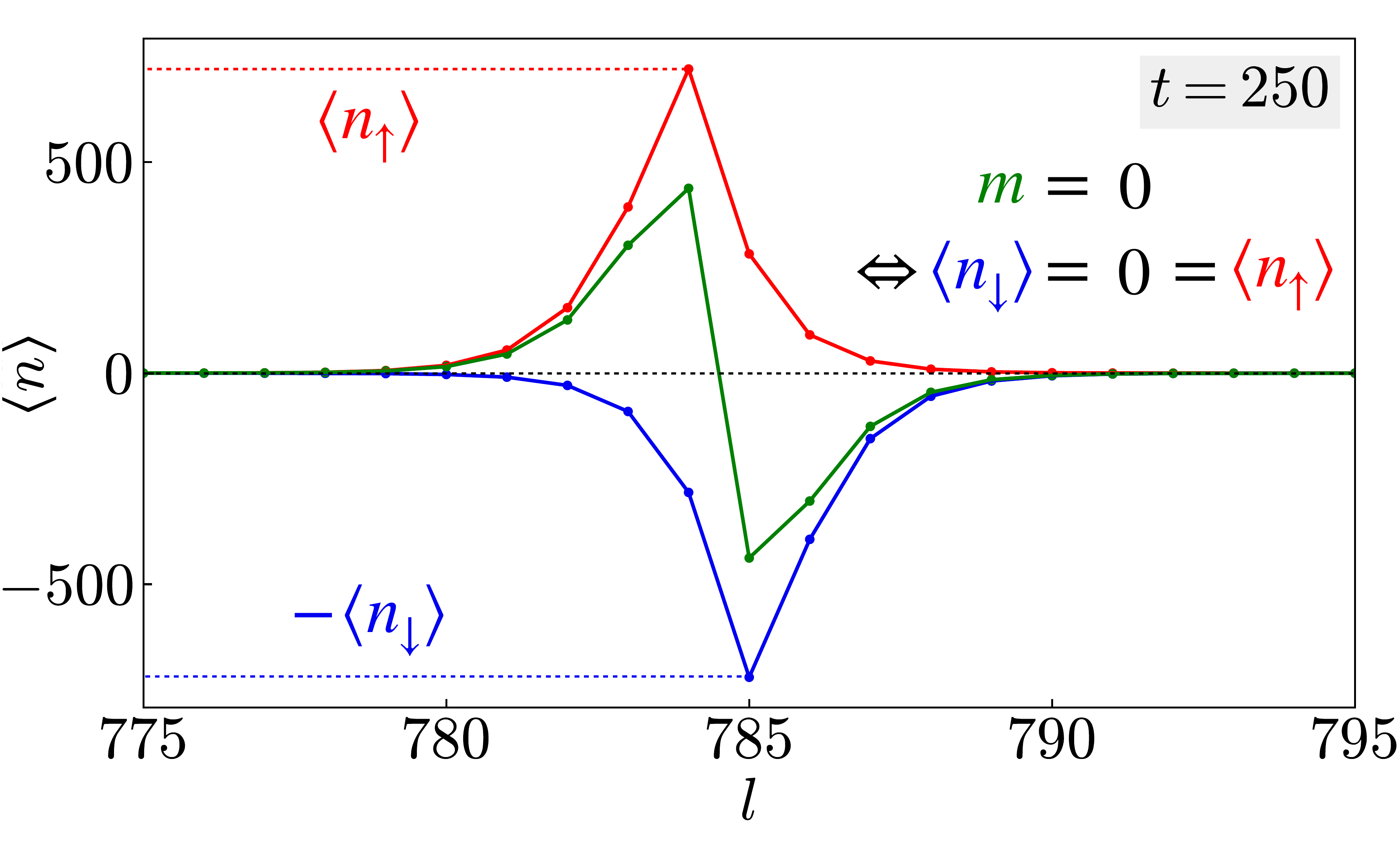}
    \label{cl:b}
  \end{subfigure}

  \vspace{-2.5em}
 
  \begin{subfigure}[t]{0.35\textwidth}
    \centering
    \caption{\raggedright}
    \includegraphics[width=\textwidth]{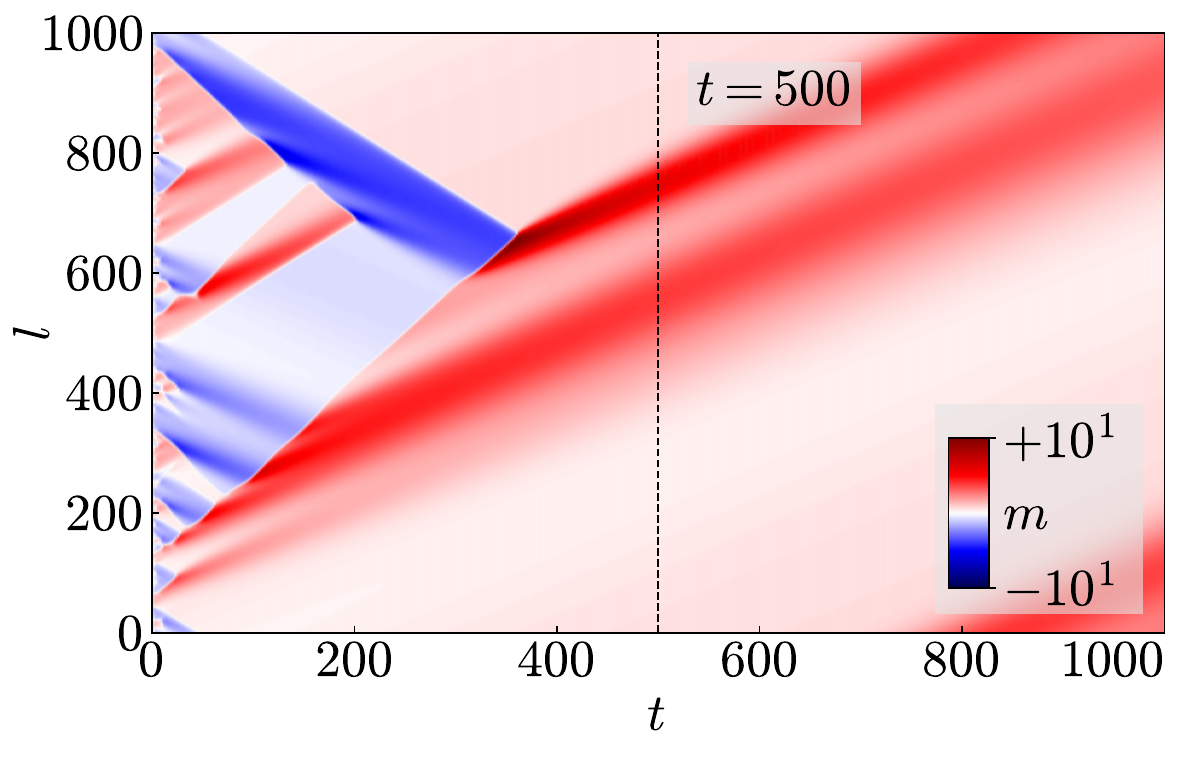}
    \label{cl:d}
  \end{subfigure} \hspace{0.3cm}
    \begin{subfigure}[t]{0.37\textwidth}
    \centering
    \caption{\raggedright}
    \includegraphics[width=\textwidth]{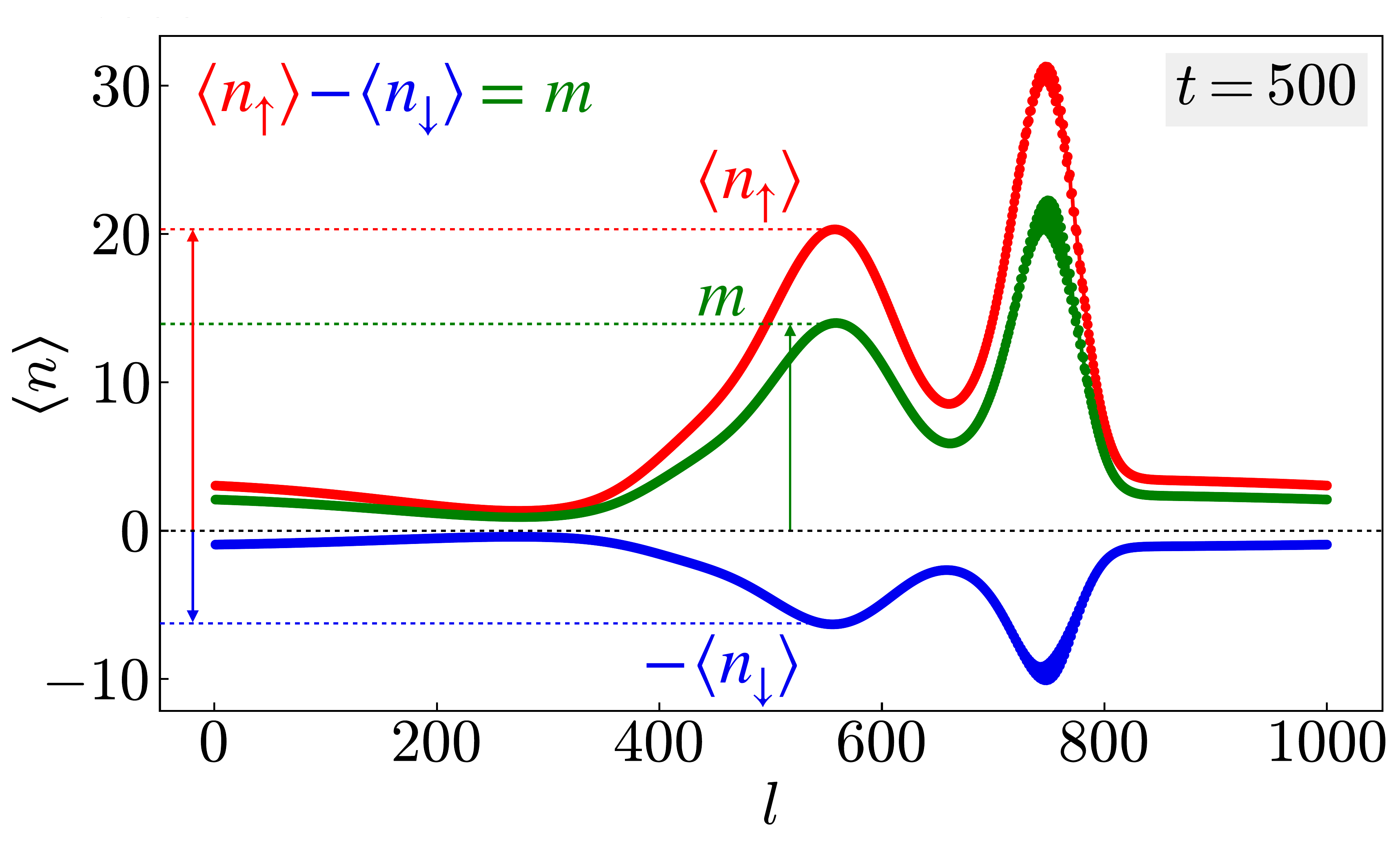}
    \label{cl:e}
  \end{subfigure}
  \vspace{-1.5em}
  \caption{\label{cl:ges}\justifying
  Dynamics of the mean-field kinetic equations of the AIM with classical particles ($L=1000$, $N=10000$, $D=1$ and $\varepsilon=0.7$) developing (a) asters at $T=0.35$ as well as (c) flocks at $T=0.45$. 
  (b),(d) Spin densities and magnetization along vertical cuts through (a) and (c), respectively, at a late time [indicated by vertical dashed lines in (a) and (c)]. In (b) we zoom into a few sites, to resolve the shape of an aster. 
  }
\vspace{-1.5em}
\end{figure*}

To investigate the effect of quantum coherences onto the dynamics,  we include the von-Neumann term $\mathcal{L}_\text{coh}[\rho] = -i[H,\rho]$ governing the coherent dynamics under the external magnetic field. 

To arrive at a kinetic equation that is also valid in presence of a magnetic field $\omega\neq 0$, we derive
\begin{align}
     \timederiv \langle n_{k\uparrow} \rangle_\mathrm{coh} &=
    -\frac{i}{\hbar} \text{tr}\left( n_{k \uparrow}[H,\rho(t)]\right) 
    =
    \frac{i}{\hbar} \langle [H, n_{k \uparrow}]\rangle\\ 
    &=-i\omega \left(\langle a_{k \uparrow}^\dagger a_{k\downarrow} \rangle - \langle a^\dagger_{k\downarrow} a_{k \uparrow} \rangle \right)
\end{align}
We observe that in the presence of the magnetic field, the on-site occupations do not decouple from the coherences anymore. Hence, when solving the dynamics numerically, we have to keep track of the full single-particle density matrix $\langle a^\dag_{l’s'}a_{ls} \rangle$. However, we suspect that terms with $l’\neq l$ are not created during the dynamics of the system and can, thus, be neglected.

 In order to arrive at a closed set of kinetic equations, similar derivations as above have to be performed for equations of motion of the quantum coherences $\langle a_{k \uparrow/\downarrow}^\dagger a_{k \downarrow/\uparrow}\rangle$, giving for the coherent part of the dynamics
\begin{align}
    \timederiv \langle a^\dagger_{k\uparrow } a_{ k\downarrow}\rangle_\mathrm{coh} &=-i\omega \left(\langle n_{k \uparrow} \rangle - \langle n_{k\downarrow} \rangle \right).
\end{align}
To arrive at a kinetic equation stemming from the dissipative hopping, we use that we neglect the coherences between neighbouring sites, giving
\begin{align}
    \begin{split}
    \timederiv \langle  a^\dagger_{k\uparrow} a_{k \downarrow}\rangle_\mathrm{H} 
    =& 
    - 2D \langle a^\dagger_{ k \uparrow} a_{k \downarrow}\rangle.
    \end{split}
    \label{D_3456}
\end{align}

For the part of the kinetic equation due to the dissipative flip, after a lengthy calculation employing Wick's theorem, cf.~\cref{app:ke-coh}, we obtain in leading order
\begin{align}
\begin{split}
    \timederiv \langle a^\dagger_{k \uparrow}  a_{k \downarrow}&\rangle_\mathrm{F} 
    = -\gamma \langle a_{k \uparrow}^\dagger  a_{k \downarrow} \rangle  \\
     +\gamma \beta^*   \langle a_{k \uparrow}^\dagger a_{k \downarrow} \rangle&\Bigl( 8\langle n_{k\uparrow} \rangle \langle n_{k\downarrow} \rangle + 4 \langle a^\dagger_{k\uparrow} a_{k\downarrow} \rangle \langle a^\dagger_{k\downarrow} a_{k\uparrow} \rangle \\
     &-6 \langle n_{k\uparrow} \rangle^2 -6 \langle n_{k\downarrow} \rangle^2 - 2\langle n_{k\uparrow} \rangle - 2\langle n_{k\downarrow} \rangle \Bigr).
\end{split}
     \label{D_12}
\end{align}
With this (and by using that  $\langle a_{k \uparrow}^\dagger a_{k \downarrow}\rangle= \langle a_{k \downarrow}^\dagger a_{k \uparrow}\rangle^*$), the kinetic equations for the dynamical variables $\{\langle n_{ks}\rangle, \langle a_{k s}^\dagger a_{k, -s}\rangle\}$ have been derived fully  and we will   investigate them  numerically in the following chapter.

\section{Quantum Flocking and Asters}\label{simulations}

In the following, in \cref{sim1,sim2,sim3}, we will first discuss the purely dissipative dynamics under vanishing external transverse magnetic field, and later, in \cref{sec:coh-spinflip}, investigate the influence of additional coherent spin flips due to an external magnetic field. 

In absence of an external magnetic field, the coherences disappear, leaving us with a classical master equation between the populations of the Fock states of the many-body system at long times. For classical (distinguishable) particles, the kinetic equations of which are
are given in \cref{eq:kineq-classical},
we expect to recover the physics of the classical active Ising model (AIM)
discussed in \cref{eq:aim-class-phases}.
In particular, we expect to approximately reproduce the AIM phase diagram  shown in \cref{mc_sim}, which we obtained by Monte-Carlo simulations (i.e., without a mean-field approximation). This will be discussed in \cref{sim1}.  
In  \cref{sim2}  we will then study the influence of the bosonic enhancement factors for indistinguishable bosons.
In \cref{sim3}, we then discuss the mean-field phase diagrams both for the classical particles as well as in case of the bosons.
Finally, in \cref{sec:coh-spinflip}, we will investigate the impact of coherent spin rotations induced by the transverse magnetic field. 

In the classical AIM, \cref{aim_intro}, the spin flipping rate is chosen as
\begin{align}
    \timederiv n_{\uparrow_F} \propto \exp\left[ \beta \left(\frac{n_\uparrow-n_\downarrow}{n_\uparrow+n_\downarrow}\right)\right],
    \label{eq:fliprate-class}
\end{align}
which deviates from the classical flipping rate $\propto \exp\left[\beta^* \left({n_\uparrow-n_\downarrow}\right)\right]$ in  \cref{eq:flip-nocoh-class} for our model.
The term in \cref{eq:fliprate-class} bounds the exponent in such a way that it remains of the order of unity.
For our quantum model, when directly numerically solving the kinetic equations derived in \cref{eq:flip-nocoh}, the  exponent in the flipping rate can become very large, which causes numerical instability.
However, we lack a straight-forward way to obtain the expectation value of the operator inverse $\langle 1/x\rangle$ in powers of $\langle x\rangle$. 
For that reason, when defining the model, we have removed the factor $1/(n_\uparrow+n_\downarrow)$. 
To make the numerics stable (and make the following simulations even possible) as well as for comparison against the classical AIM, we introduce this prefactor by hand.
This means that we adapt
$\beta^*$ according to \cref{eq:beta-star}
and, in that way, reintroduce the inverse density dependence in the kinetic equations.

\begin{figure*}[t]
  \begin{subfigure}[t]{0.35\textwidth}
    \centering
    \caption{\raggedright}
    \includegraphics[width=\textwidth]{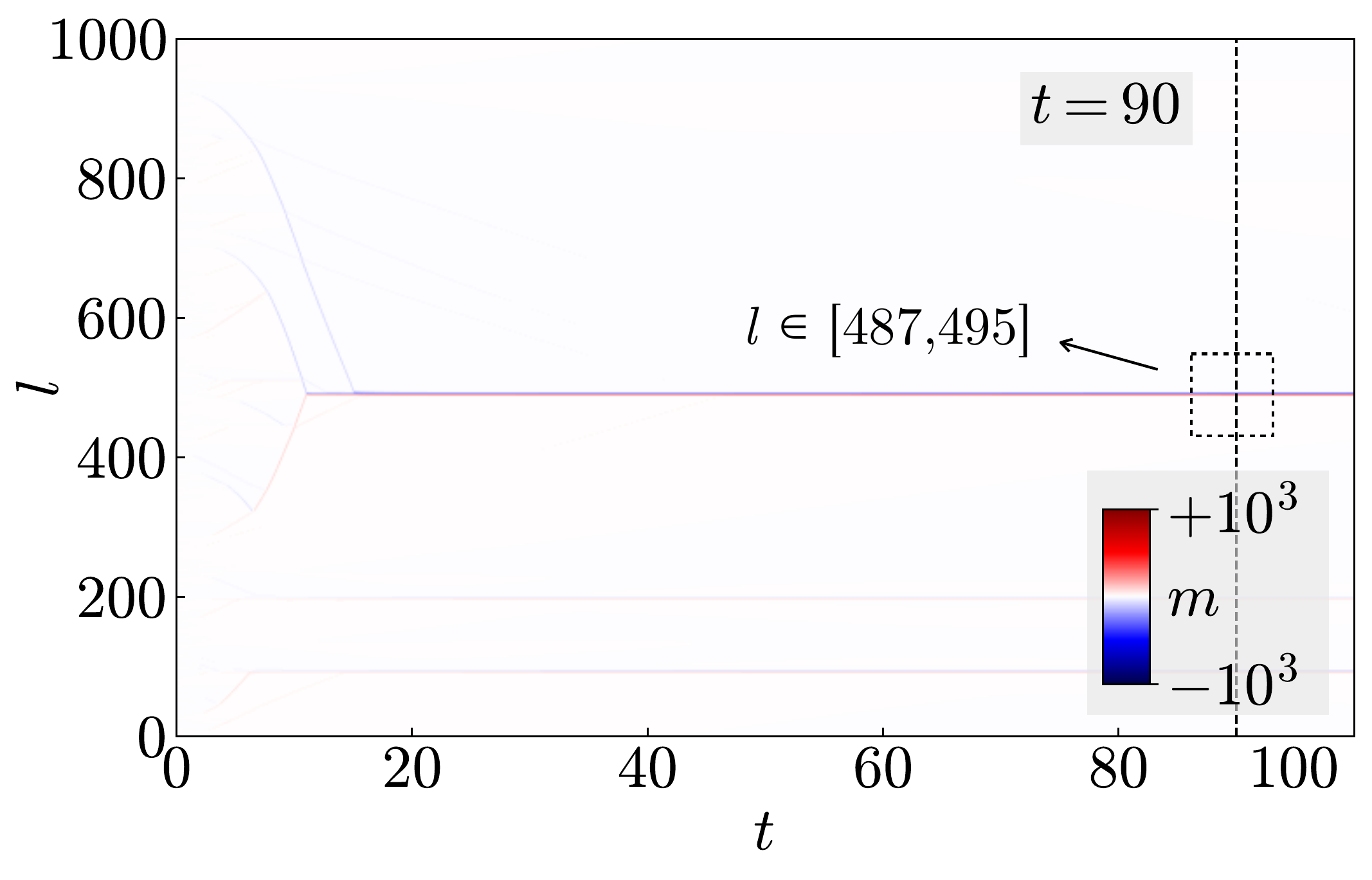}
    \label{qm:a}
  \end{subfigure}
    \begin{subfigure}[t]{0.38\textwidth}
    \centering
    \caption{\raggedright}
    \includegraphics[width=\textwidth]{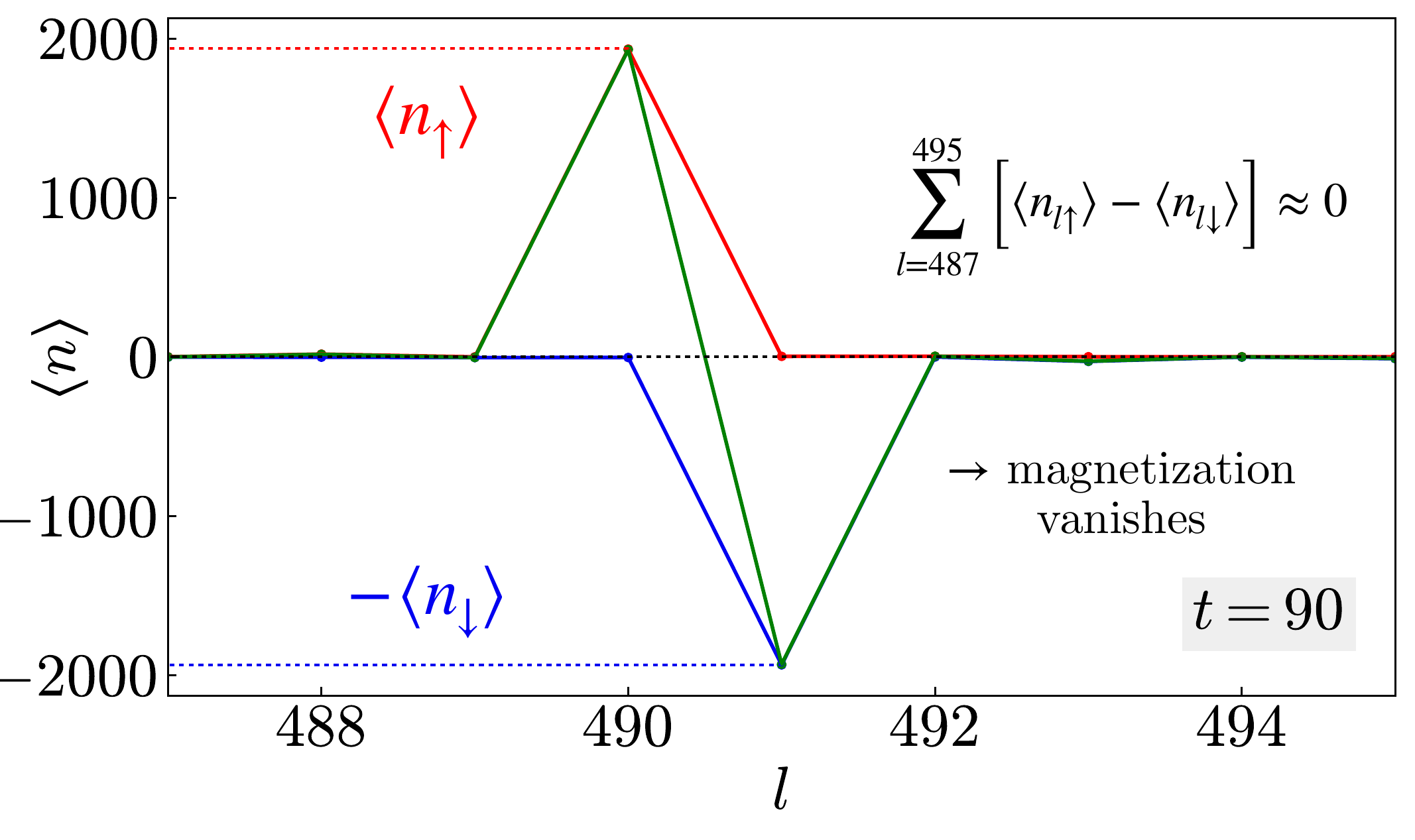}
    \label{qm:b}
  \end{subfigure}

  \vspace{-2.5em}

  \begin{subfigure}[t]{0.35\textwidth}
    \centering
    \caption{\raggedright}
    \includegraphics[width=\textwidth]{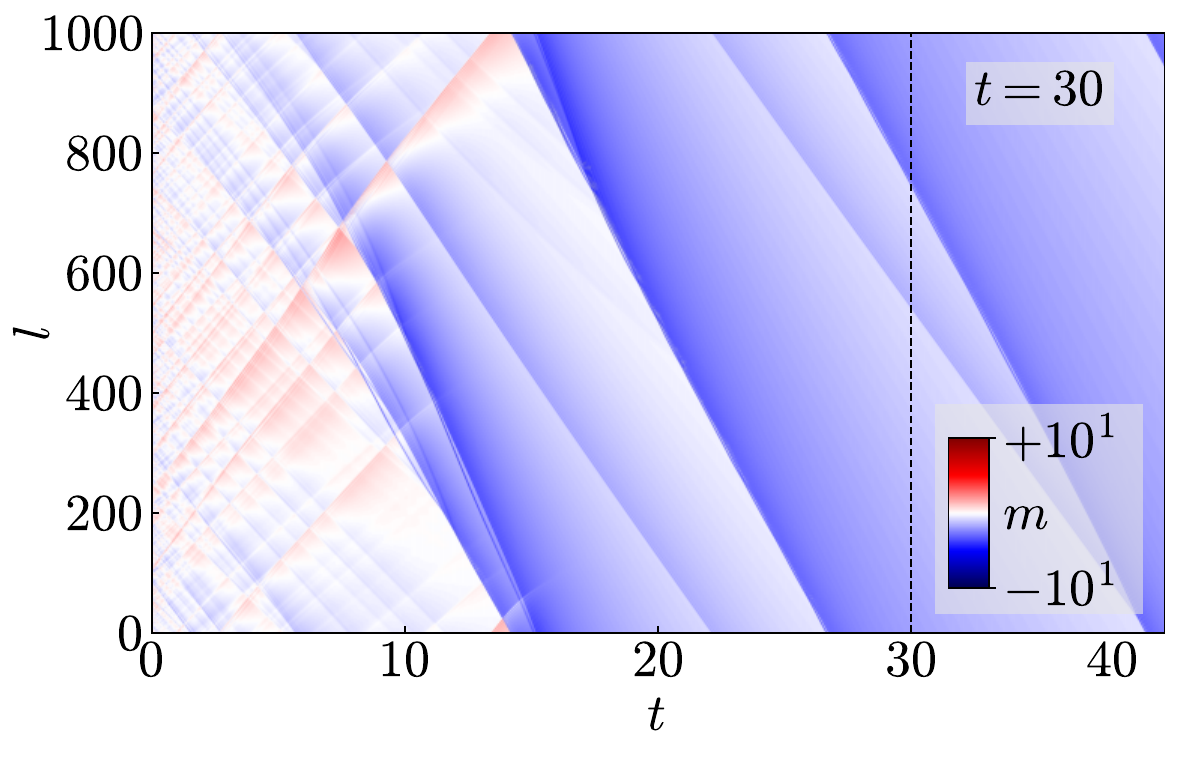}
    \label{qm:d}
  \end{subfigure}
    \begin{subfigure}[t]{0.35\textwidth}
    \centering
    \caption{\raggedright}
    \includegraphics[width=\textwidth]{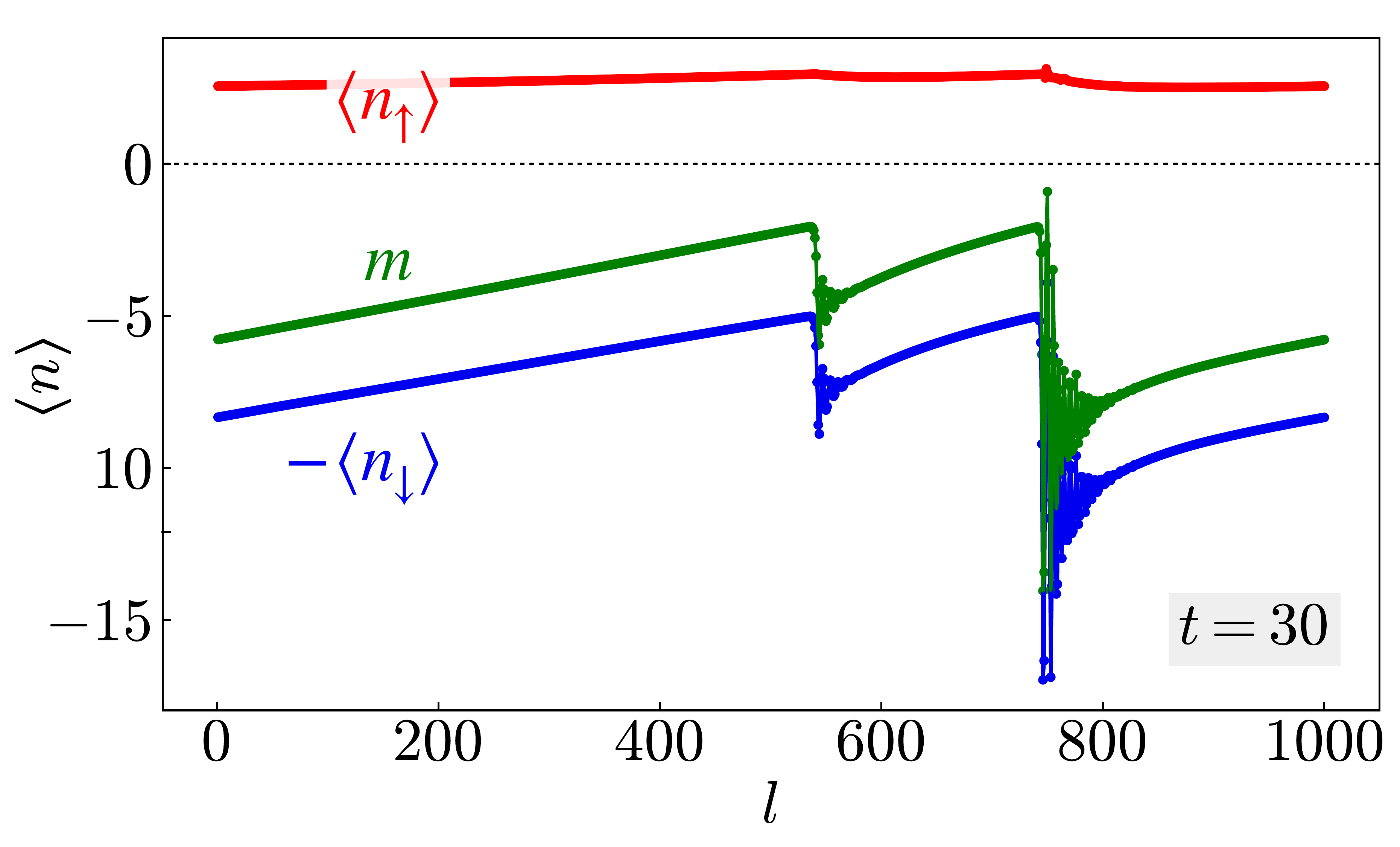}
    \label{qm:e}
  \end{subfigure}
  \vspace{-1.5em}
  \caption{\label{qm:ges}\justifying
    Dynamics of a large quantum AIM of ideal bosons ($L=1000$, $N=10000$ and $\varepsilon=0.7$) developing (a) asters at $D=0.2$ and $T=7$ as well as (c) flocks at $D=3$, $T=10$.
    (b),(d) 
    Absolute magnetization of each lattice site at a late time of figures (a) and (c), respectively [indicated by vertical dashed lines in (a) and (c)]. The aster phase evolves differently when compared to the classical model, with multiple neighbouring peaks of opposite magnetization.
    Additionally, the flock appears sharply localized at its edge instead of the continuous curve found in \cref{cl:ges}.
  }
\end{figure*}

\begin{figure*}[t]
  \centering
  \begin{subfigure}[t]{0.32\textwidth}
    \centering
    \caption{\raggedright\@$L=20$}
    \includegraphics[width=\textwidth]{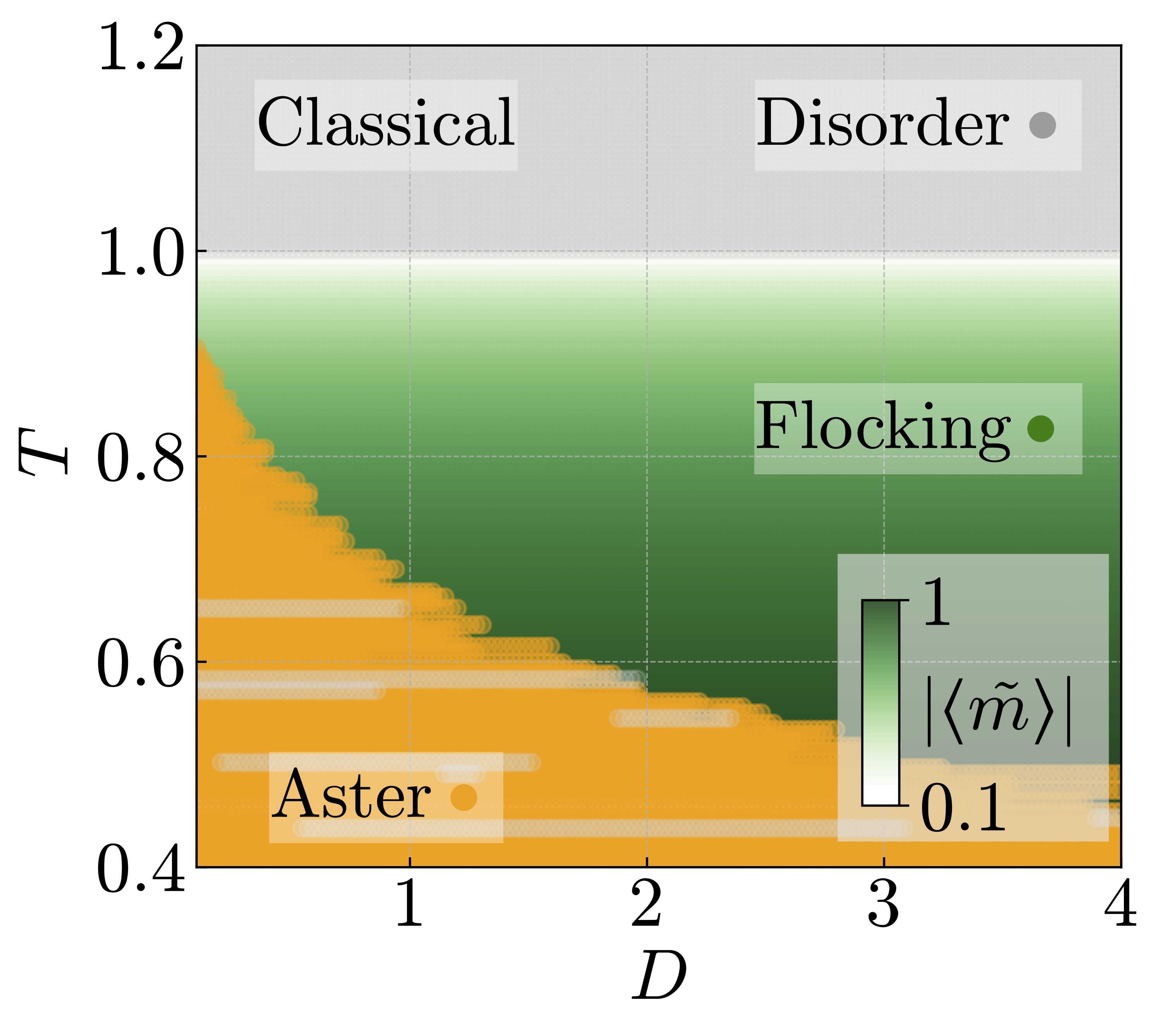}
    \label{pd:a}
  \end{subfigure}\hfill
    \begin{subfigure}[t]{0.32\textwidth}
    \centering
    \caption{\raggedright\@$L=100$}
    \includegraphics[width=\textwidth]{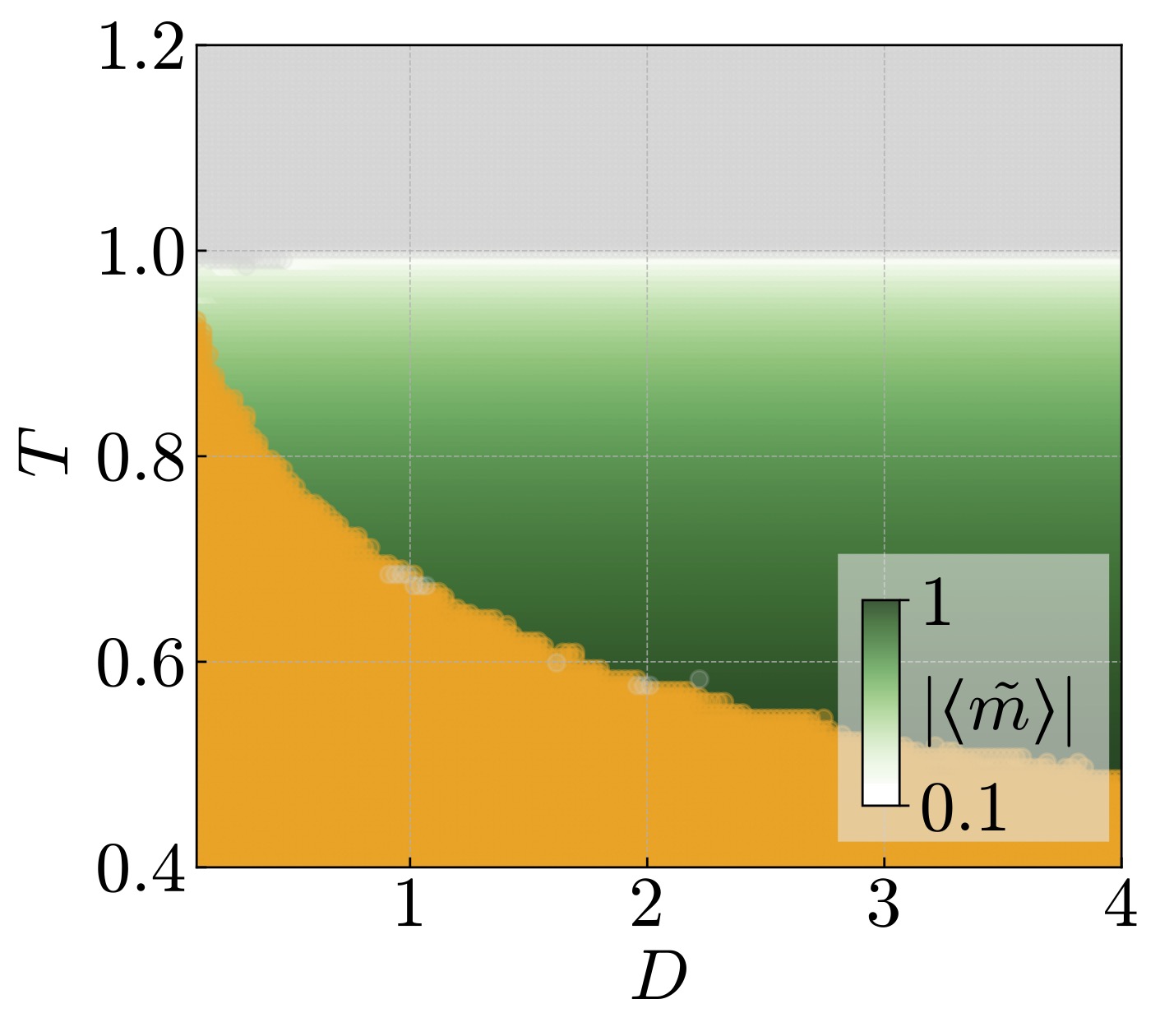}
    \label{pd:b}
  \end{subfigure}\hfill
  \begin{subfigure}[t]{0.32\textwidth}
    \centering
    \caption{\raggedright\@$L=100$}
    \includegraphics[width=\textwidth]{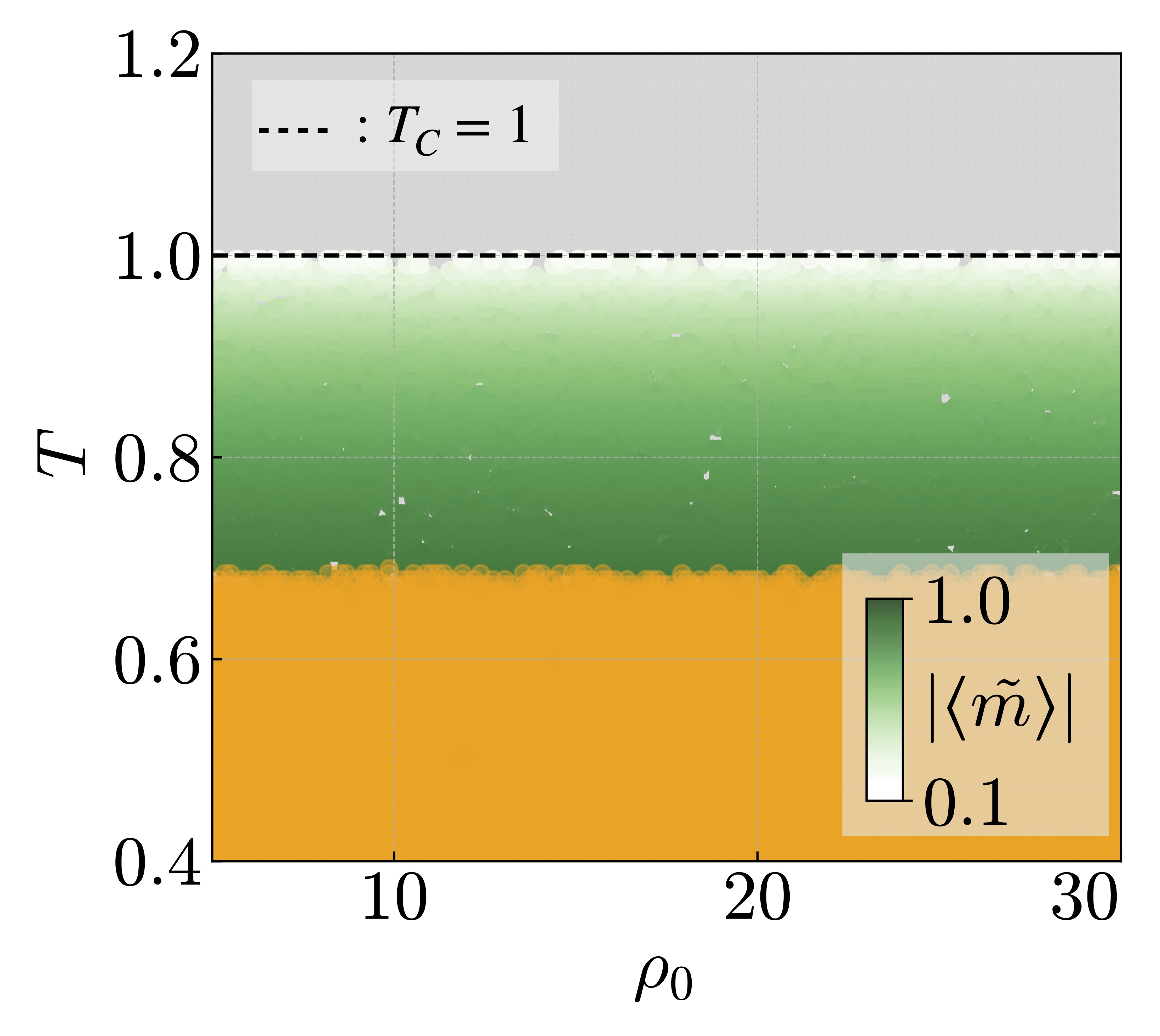}
    \label{pd:c}
  \end{subfigure}

  \vspace{-1.5em}
 
  \begin{subfigure}[t]{0.32\textwidth}
    \centering
    \caption{\raggedright\@$L=100$, $e^\beta=T_1(\beta)$}
    \includegraphics[width=\textwidth]{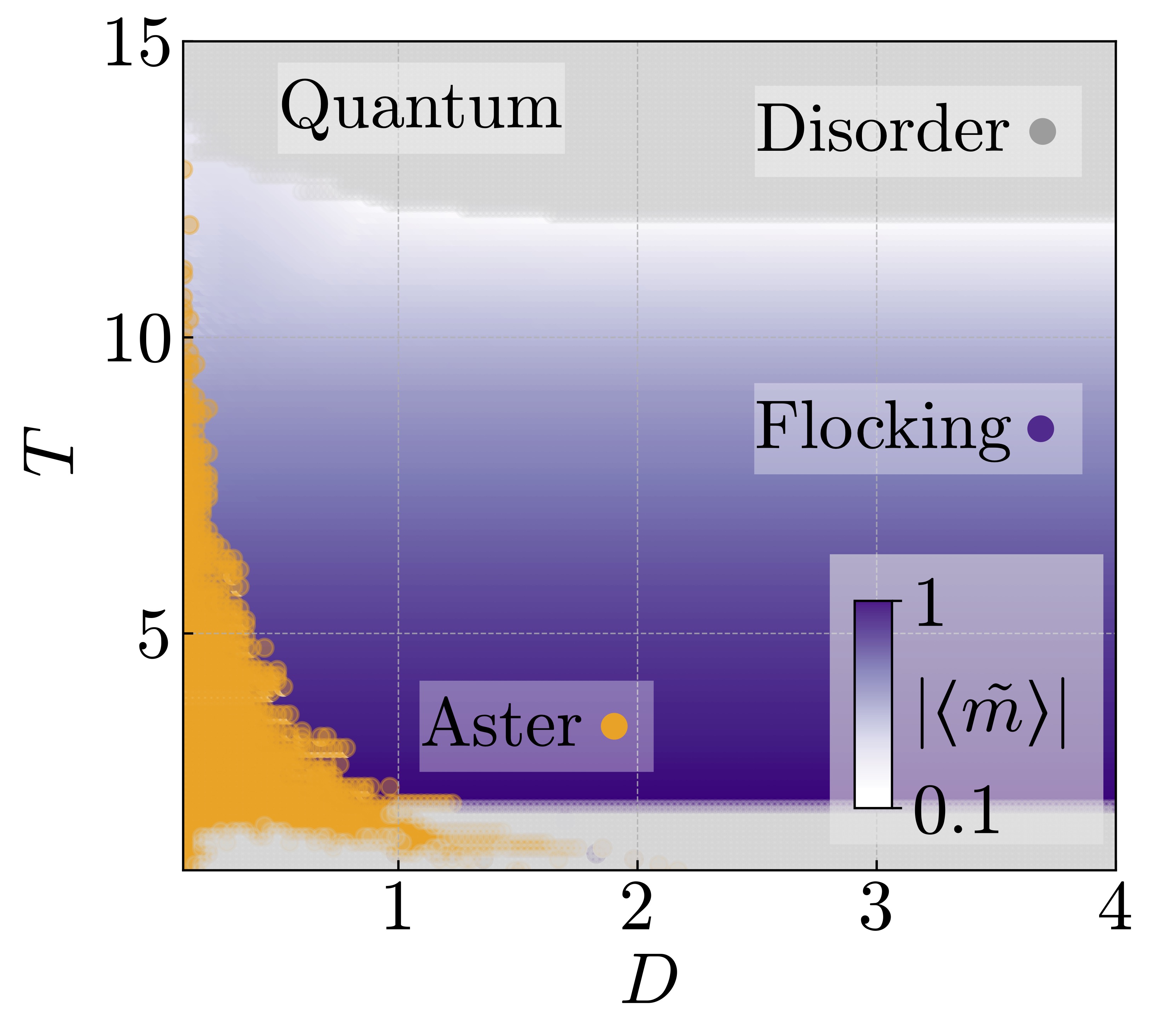}
    \label{pd:d}
  \end{subfigure}\hfill
    \begin{subfigure}[t]{0.32\textwidth}
    \centering
    \caption{\raggedright\@$L=100$, $e^\beta=T_4(\beta)$}
    \includegraphics[width=\textwidth]{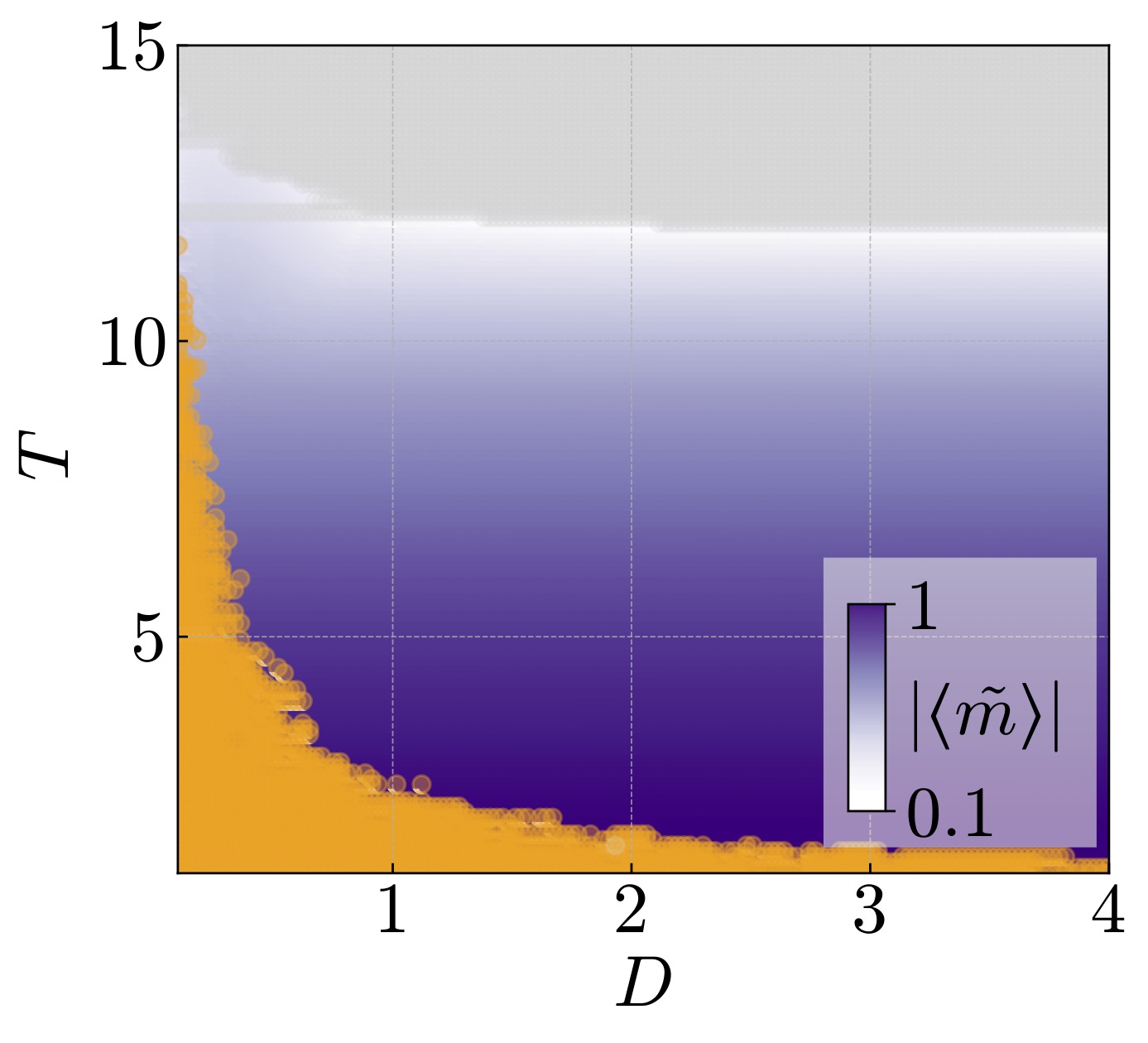}
    \label{pd:e}
  \end{subfigure}\hfill
  \begin{subfigure}[t]{0.32\textwidth}
    \centering
    \caption{\raggedright\@$L=100$, $e^\beta=T_1(\beta)$}
    \includegraphics[width=\textwidth]{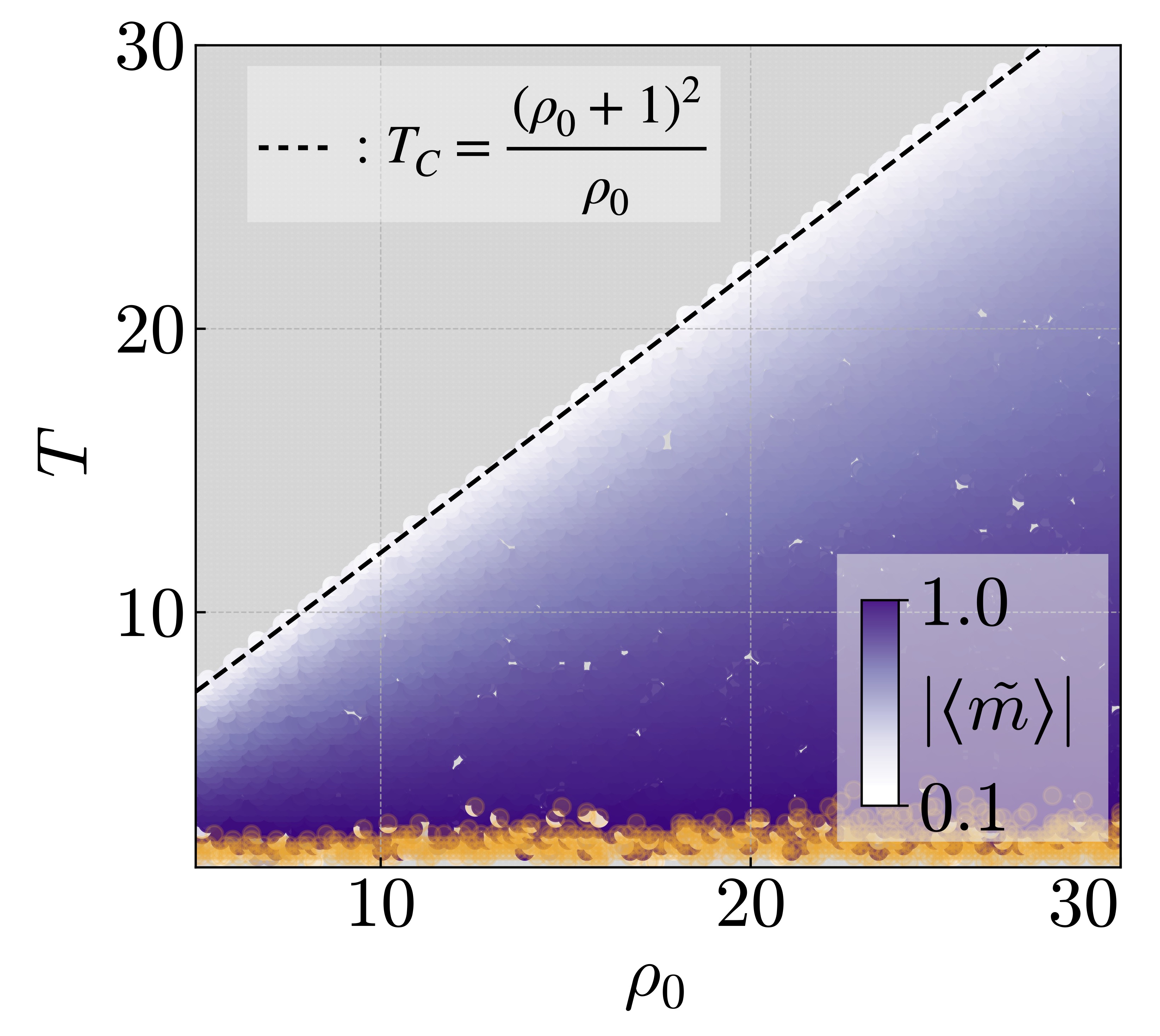}
    \label{pd:f}
  \end{subfigure}
  \vspace{-1.5em}
  \caption{\label{pd:ges}\justifying
    Different phase diagrams of the classical- and quantum AIM without transverse magnetic field.\@
    Flocking phases are signaled by a finite mean magnetization, as indicated by the color code, whereas the presence of asters is marked by yellow dots. In the absence of both, the disordered phase is marked by grey shading. 
  (a),(b),(d),(e) Phase diagram in the plane spanned by $T$ and $D$ for two different system sizes, $\rho_0 =10$  and $\varepsilon=0.7$. 
  (c),(f) Phase diagram in the plane spanned by $T$ and $\rho$ for $D=1$ and $\varepsilon=0.7$.
  In
  (d),(f) the mean-field equation is used with a high-temperature Taylor expansion up to first order and  in
  (e) up to fourth order to compare both approaches. In both (c) and (f), the dashed lines correspond to mean-field estimates of the transition temperatures from the disordered to the flocking state.}
\vspace{-1.5em}
\end{figure*}

\subsection{Classical system}\label{sim1}
 Before turning to the quantum system characterized by bosonic enhancement and possibly also a coherent transverse field, let us first compare the dynamics under the classical master equation  for the AIM as in  \cref{mc_sim} and Ref.~\cite{classical_flocking} with the results of the kinetic equation for the classical model, \cref{eq:kineq-classical}. 
In \cref{cl:ges}, we show the dynamics in both nontrivial phases, the aster and the flocking phase. 
We plot the site-resolved mean magnetization $\langle m_l \rangle$ versus position $l$ and time $t$ in \cref{cl:a,cl:d} for $N=10000$ particles on $L=1000$ sites, with randomly chosen initial conditions. 
\cref{cl:a} depicts a typical example of the dynamics in the classical aster phase, where we can observe the formation of several asters, whereas \cref{cl:d} shows a typical evolution in the flocking phase, where the formation of a large flock of particles with positive magnetization can be observed that is moving in positive $l$ direction.

We can distinguish different stages in the evolution. In a first stage, we find in both regimes the formation of small local flocks, with the spin symmetry being broken spontaneously. In a second stage, flocks of opposite polarization collide. In the flocking regime, these collisions lead to the merging of flocks, with the larger flock absorbing the smaller one by converting its magnetisation via the spin alignment process. In the aster phase, the outcome of the collision of two opposing flocks depends on their respective sizes: The collision of two small local flocks of different size, can also lead to merging processes  as in the flocking phase. However, when two larger flocks collide, an a stationary aster is formed. The so formed asters can then grow by absorbing small flocks. Finally, in a third stage a steady state is reached. In the aster phase, it consists of randomly placed asters of different size, whereas in the flocking phase macroscopic flocks have formed that spontaneously break the spin symmetry and all move in the same direction.

In \cref{cl:b}, we depict the spatial distribution of the spin-densities and the resulting magnetization for a single aster at a late time, indicated by the dashed vertical line in (a). Both density and magnetization are localized strongly at the aster and rapidly approach zero when moving away from the center of the aster. In \cref{cl:d}, the spin densities and the magnetization are shown for a late time in the flocking phase [indicated by the dashed line in \cref{cl:d}]. We clearly see that the spin-symmetry is broken globally by the formation of two large overlapping flocks, comoving in positive $l$ direction [according to \cref{cl:d}]. The density distribution of each of the flocks is roughly symmetric around their respective instantaneous center.

\subsection{Ideal bosons}\label{sim2}

Having now investigated the  dynamics of the kinetic equations for classical particles, we turn to  the discussion of the dynamics for indistinguishable bosons.
Generally, the same phases can be observed as shown in \cref{qm:ges}. 
In \cref{qm:a}, we show the dynamics in the aster phase. Here, the system behaves rather similarly compared to the classical case. 
In \cref{qm:d}, we show the typical dynamics in the flocking phase. Remarkably, we find that the time needed for the formation of a flock is, roughly, ten times smaller that in the classical model. This factor can, intuitively, be explained by noting that the bosonic enhancement leads to a speed-up, which on average is proportional to the mean density $\rho_0=10$.

We also observe that instead of the formation of several smaller asters, one large aster is formed. This suggests that, as a result of the quantum statistics, the stabilization of an aster requires a larger density.  
This can be seen in \cref{qm:b}, where a snapshot of both the density and the absolute value of the magnetization for a late time is shown.
Whereas, in the aster phase, the profile looks roughly similar to the classical case, shown in \cref{cl:b}, in the flocking phase, for which a similar snapshot is shown in \cref{cl:e}, we find noticeable difference between the quantum and the classical model. 
Instead of a mostly symmetric shape of the classical flock, the quantum flock seems to possess a sharp front and then  drops smoothly after initial density oscillations.
We clearly observe how 
the profile slowly decays before the magnetization reaches another peak as the next flock appears.

\begin{figure*}[t]
  \centering
  \begin{subfigure}[t]{0.32\textwidth}
    \centering
    \caption{\raggedright}
    \includegraphics[width=\textwidth]{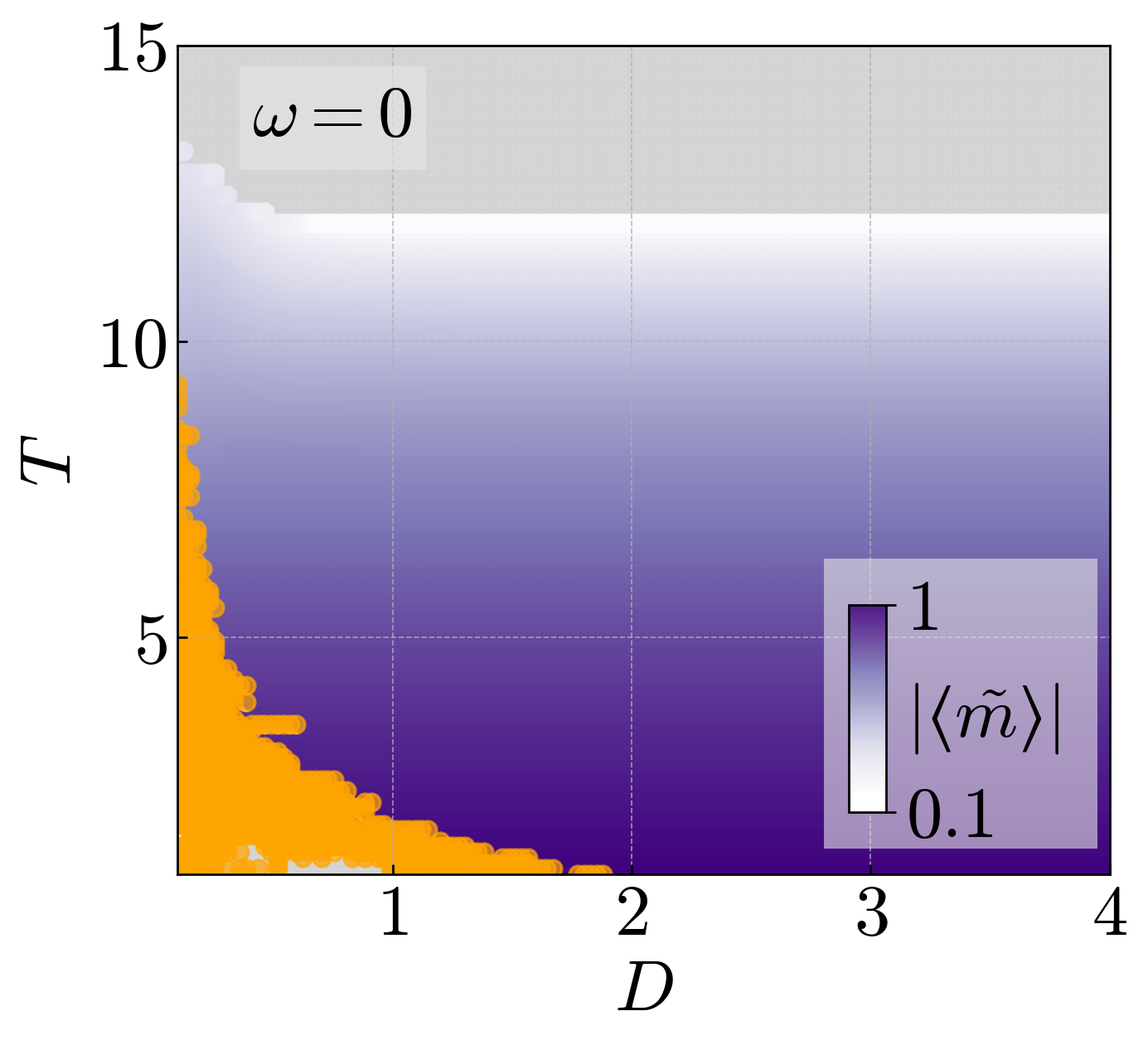}
    \label{pd_H:a}
  \end{subfigure}\hfill
    \begin{subfigure}[t]{0.32\textwidth}
    \centering
    \caption{\raggedright}
    \includegraphics[width=\textwidth]{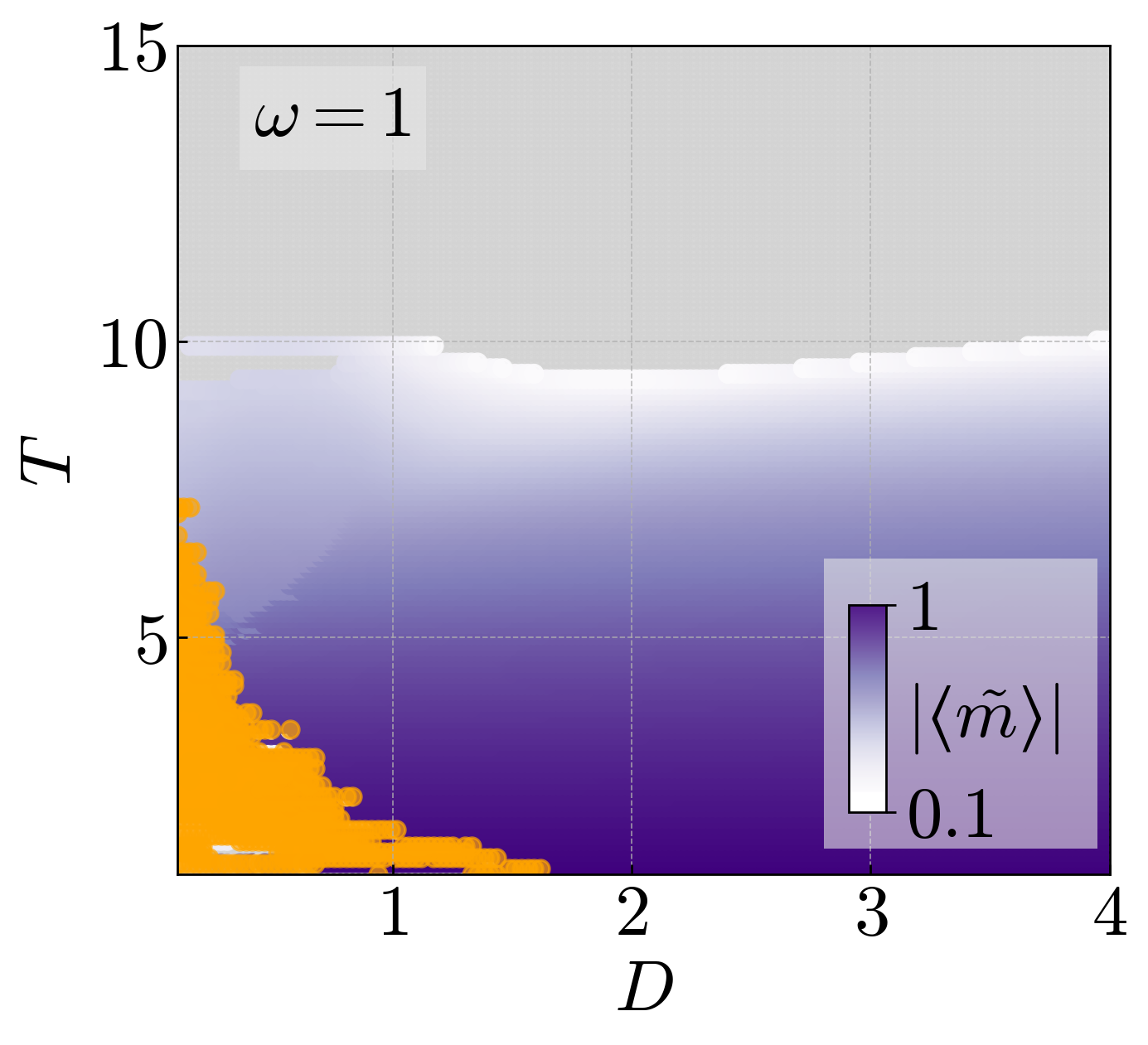}
    \label{pd_H:b}
  \end{subfigure}\hfill
  \begin{subfigure}[t]{0.32\textwidth}
    \centering
    \caption{\raggedright}
    \includegraphics[width=\textwidth]{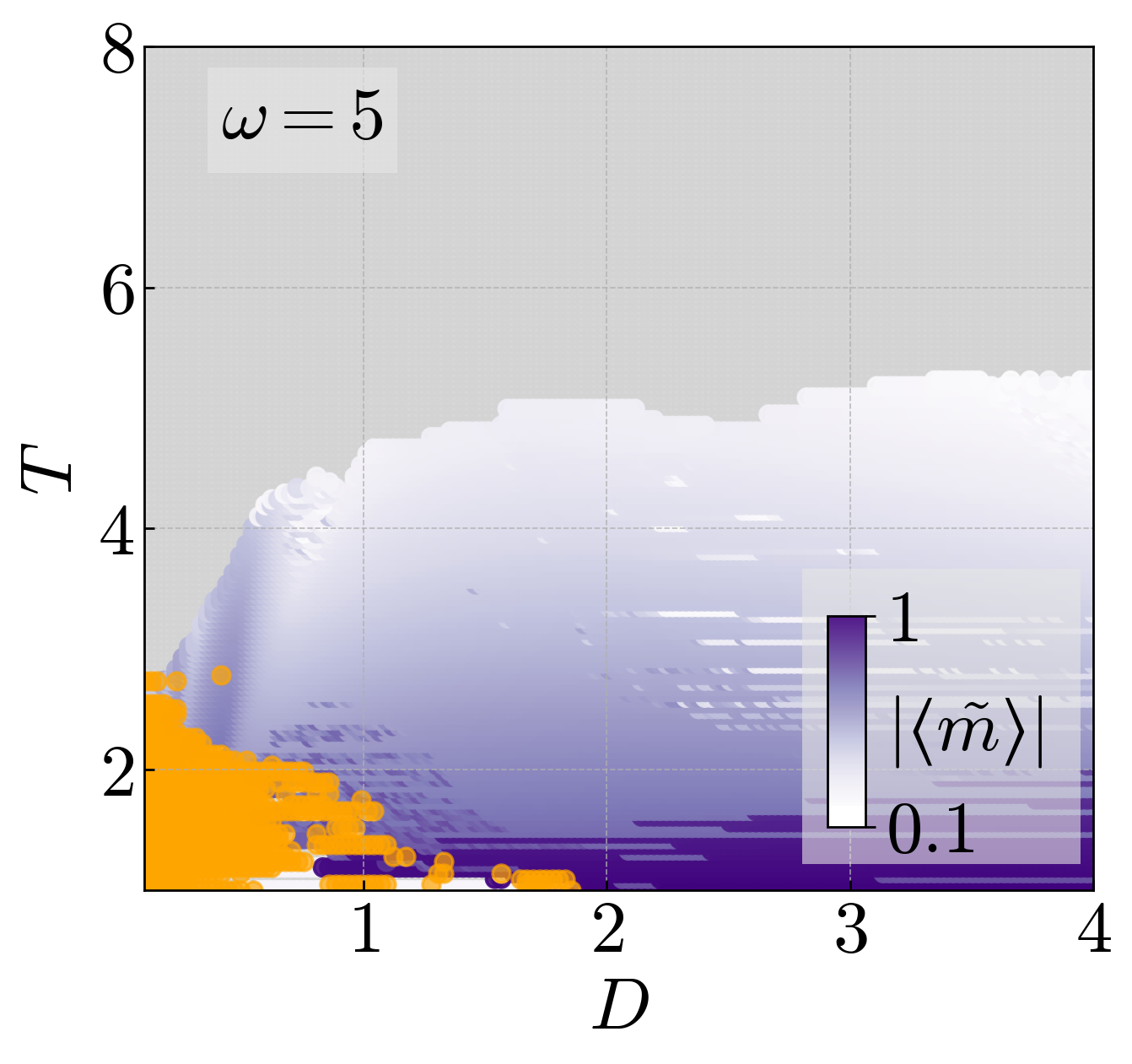}
    \label{pd_H:c}
  \end{subfigure}

  \vspace{-2.5em}
 
  \begin{subfigure}[t]{0.32\textwidth}
    \centering
    \caption{\raggedright}
    \includegraphics[width=\textwidth]{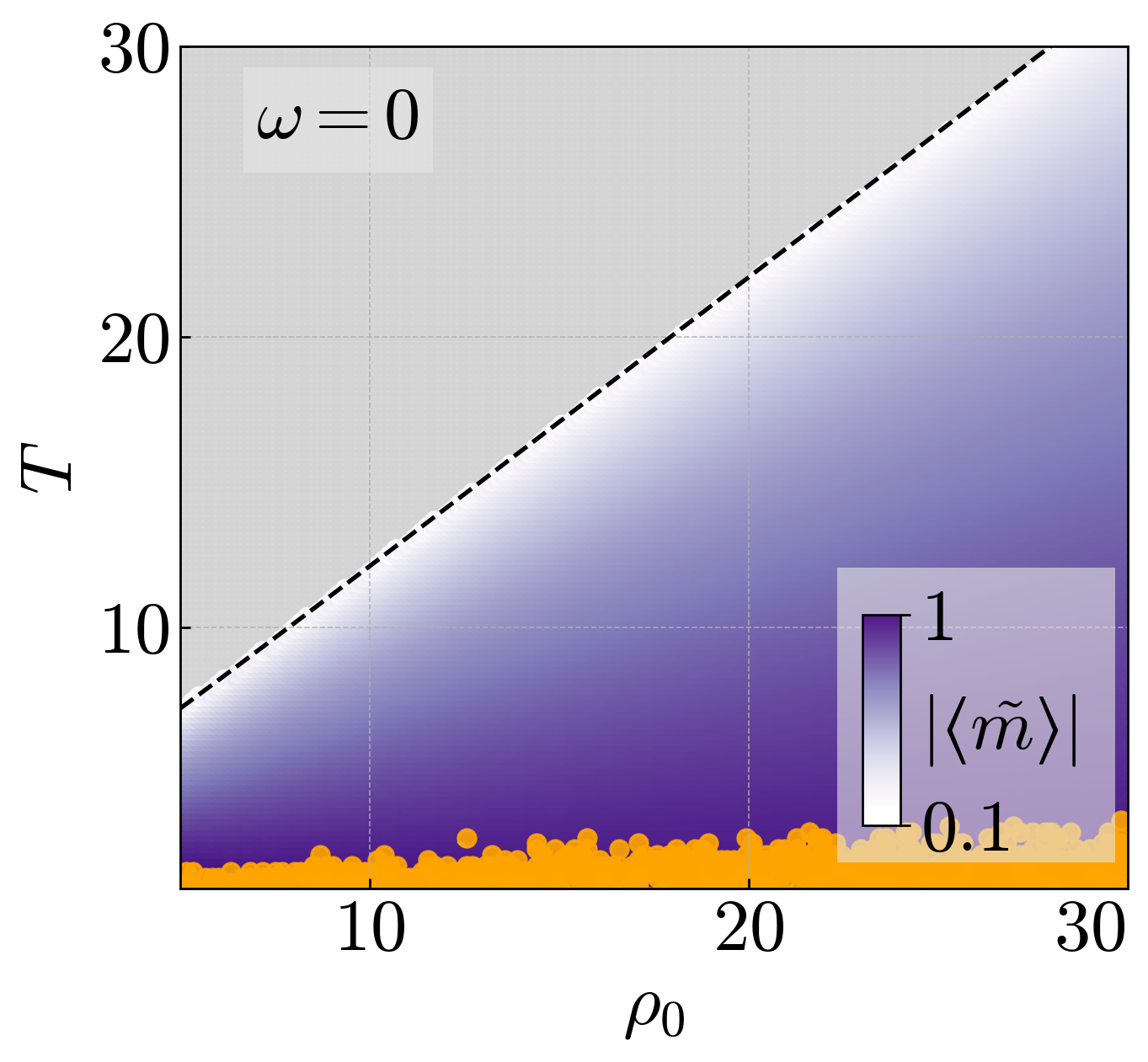}
    \label{pd_H:d}
  \end{subfigure}\hfill
    \begin{subfigure}[t]{0.32\textwidth}
    \centering
    \caption{\raggedright}
    \includegraphics[width=\textwidth]{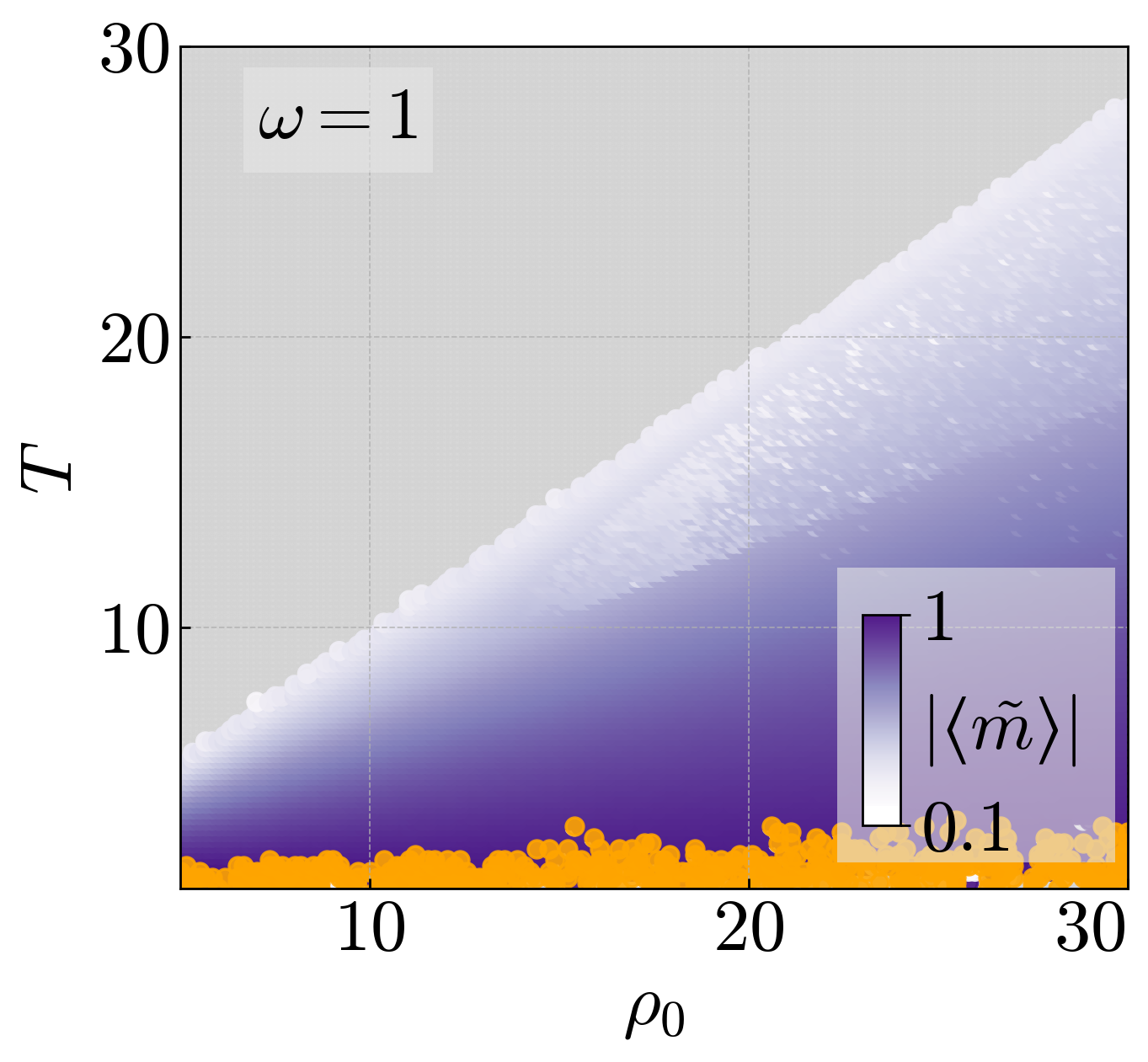}
    \label{pd_H:e}
  \end{subfigure}\hfill
  \begin{subfigure}[t]{0.32\textwidth}
    \centering
    \caption{\raggedright}
    \includegraphics[width=\textwidth]{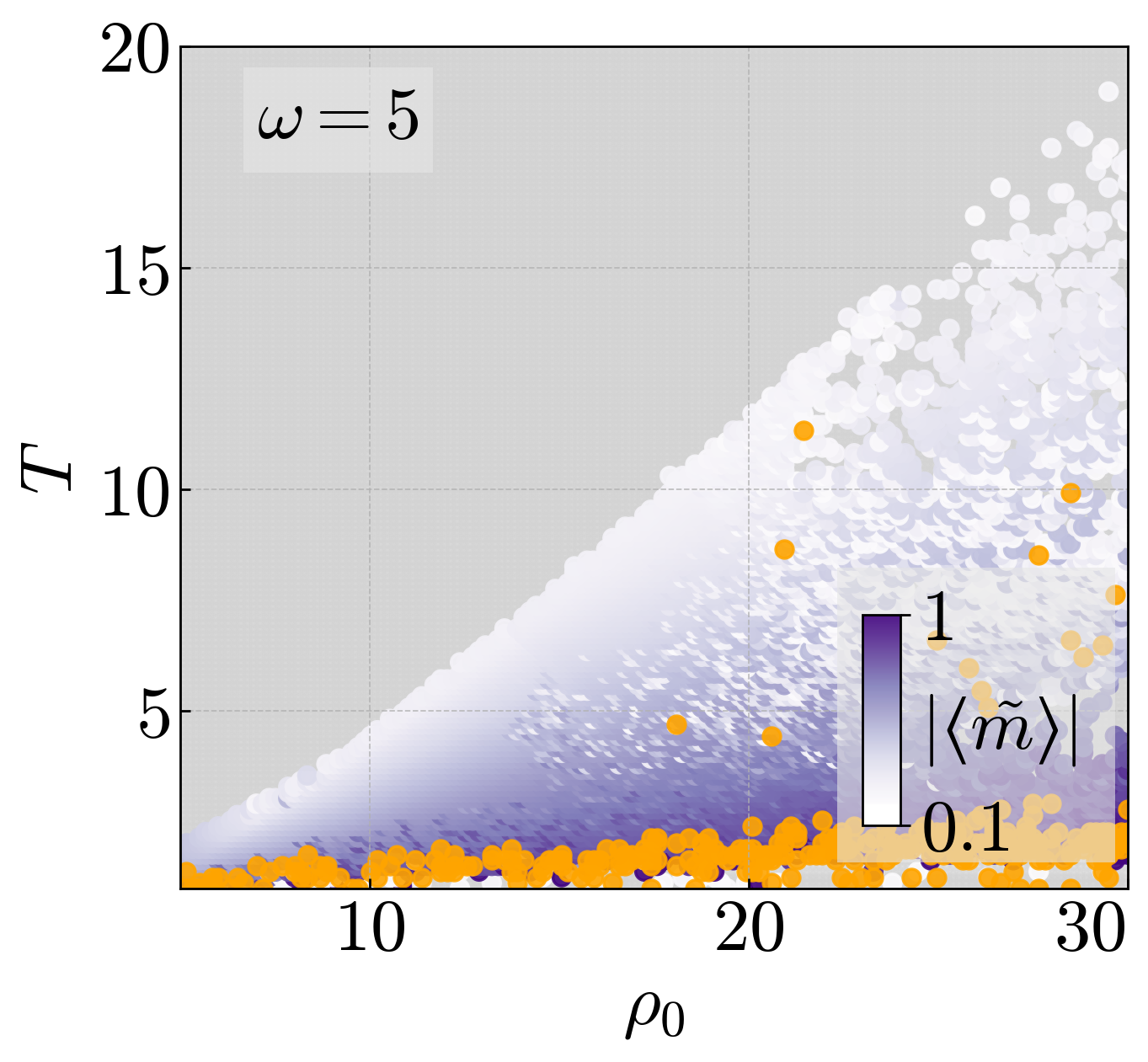}
    \label{pd_H:f}
  \end{subfigure}
  \vspace{-1.5em}
  \caption{\label{pd_H:ges}\justifying
    Phase diagrams of the quantum AIM under the influence of an external Hamiltonian with varying frequency $\omega$. 
  To avoid excess computational work, we show the phase diagrams for system size $L=20$ and again a first order temperature expansion is used. (a),(b),(c)  $T-D$ diagrams for $N=200$ and $\varepsilon=0.7$. (d),(e),(f) $T-\rho_0$ diagram for $D=1$ and $\varepsilon=0.7$. 
  We vary $\rho_0$ by changing $N\in \left[100,600\right]$.}
\vspace{-1.5em}
\end{figure*}
 
\subsection{Phase diagrams}\label{sim3}
Following this qualitative discussion of single trajectories, let us now map out the system's phase diagram. 
For that purpose, we use the same criteria that we describe in \cref{aim_intro} to also detect the different phases of the quantum active Ising model.


In \cref{pd:ges}, we show phase diagrams of both the classical- and the quantum model as they are obtained from the kinetic equations. Before turning to the quantum case, we can first observe that the mean-field phase diagram computed for the classical model in Fig.~\ref{pd:b} closely resembles that shown in Fig.~\ref{mc:ges}, which was obtained using quasi-exact Monte-Carlo simulations. This shows that all three phases can be identified clearly in mean-field theory. It is, moreover, interesting to observe that simulations for a rather small system of only 20 sites provide a phase diagram (Fig.~\ref{pd:a}) that hardly differs from the one computed for 100 sites (Fig.~\ref{pd:b}).



Turning to the quantum model, \cref{eq:n-mean-compl}, let us first verify the validity of the high-temperature expansion in \cref{en-main,nen-main}. We can see that the phase diagram in leading (first) order, shown in Fig.~\ref{pd:d} is almost identical to the one obtained in fourth order \cref{pd:e}, except for a slight shift of the boundary between aster and flocking phase at very low temperature. This suggests that the leading-order phase diagrams are already very close to the (mean-field) phase diagram of the exact model.

Comparing the phase diagrams Figs.~\ref{pd:e} and \ref{pd:b}, computed for the quantum- and the classical model, respectively, we can clearly see that the ordered flocking and aster phase, which are dominated by collective non-equilibrium processes, are favoured in the quantum model. For the given parameters, they remain stable for temperatures that are an order of magnitude larger compared to the corresponding phases of the classical model. This striking effect is a consequence of the bosonic quantum statistics. Namely, it is related to bosonic enhancement, which is also responsible for collective processes, such as Bose-Einstein condensation in equilibrium and lasing far away from equilibrium.

The impact of quantum statistics becomes relevant in the quantum degenerate regime, when the number of particles becomes comparable to or even larger than the number of accessible single-particle states. In our system, where the kinetic energy of the particles, as it would show up in the form of coherent tunneling processes, is assumed to be negligible and neglected, all sites are accessible. Thus, quantum statistics matters for densities $\rho_0\gtrsim1$. This is precisely the regime, where bosonic enhancement matters.

In order to study the impact of bosonic enhancement, we also plot phase diagrams for the classical- and the quantum model in the temperature-density plane in Figs.~\ref{pd:c} and \ref{pd:f}, respectively. While we do not observe a density dependence for the classical mean-field phase diagram, one can clearly see that the critical temperature, below which flocking occurs, increases linearly with the density $\rho_0$ in the quantum model. 

Let us investigate the impact of the statistical enhancement on the collective behaviour by examining the critical temperature for flocking analytically. 
In the classical system this was done \cite{classical_flocking} by assuming a homogeneous density and magnetization in the steady state, i.e.,
$\langle {n}_{l\uparrow} \rangle \approx \langle {n}_\uparrow \rangle$, and  
$\langle {n}_{l\downarrow} \rangle \approx \langle {n}_\downarrow \rangle.
$
With this, one derives corresponding equations of motion of the mean magnetization
$\langle {m} \rangle = \langle {n}_\uparrow \rangle - \langle {n}_\downarrow \rangle,$  
and the mean density
$ {\rho}_0 = \langle {n}_\uparrow \rangle + \langle {n}_\downarrow \rangle.$
By expanding the exponential in the flip rate up to third order in $\beta$, one then obtains a non-linear equation \cite{classical_flocking} for the stationary solution,
\begin{align}
    \timederiv \langle {m}\rangle&=2\gamma \langle m\rangle(\beta-1) -\gamma \left(\beta^2 -   \frac{\beta^3}{3}\right) \frac{\langle m\rangle^3}{{\rho}_0^2}\overset{!}{=}0.
\end{align}
This equation has nontrivial solutions $\langle m \rangle \neq 0 $ only if
\begin{align}
     T&< 1,
\end{align}
i.e.~$\beta>1$.
This result is well established \cite{Solon2015}. It implies that the critical temperature below which flocking can occur in mean-field theory is independent of the system's density \cite{Solon2015}. We show this critical temperature in \cref{pd:c} and observe good agreement with the numerics.

By a similar procedure, cf.~\cref{appendix_d}, we can quantify the impact of the density-dependent bosonic enhancement factors on the critical temperature.
We find the modified equations of motion
\begin{align}
    \timederiv \langle {m}\rangle&=2\gamma \left[\langle m\rangle \left(2\beta -1\right) + \beta \frac{\langle m\rangle}{{\rho}_0}\left(1+{\rho}_0^2-\langle m\rangle^2 \right)\right]\overset{!}{=}0.
\end{align}
Note that in contrast to the classical case, here we just expand to first order in $\beta$ since nonlinearities already occur in this order in the quantum model.
This equation has solutions with nonzero order parameter $\langle m \rangle$ for
\begin{align}
     T&< \frac{(1+\rho_0)^2}{ \rho_0}.
     \label{eq:T_qm}
\end{align}
At high densities,  the critical temperature increases approximately linearly with the density, which accurately descibes the numerical behaviour observed in \cref{pd:f}. Note that \cref{eq:T_qm} only holds in the high-density limit and does not recover the classical critical temperature for $\rho_0 \to 0$. In order to find a consistent expression that also recovers the classical limit, one would have to expand the bosonic mean-field equations up to third order.



The observation that bosonic quantum statistics enhances and stabilizes collective effects is a main finding of this paper. Note that this is a novel observation adding to the findings in 
Ref.~\cite{quantum_flocking}, where quantum flocks have been found for hard-core bosons. Due to the hard-core constraint, in that model the effects of bosonic enhancement on quantum flocking are not present. The stabilization of collective behaviour by the bosonic quantum statistics is contrary to the suppression of collective processes by quantum fluctuations that are induced by a transverse magnetic field and that counteract the spin-alignment \cite{quantum_flocking}.  
In the next section, we study our quantum model in the regime where both effects are present and compete.


\subsection{Competition of quantum statistics and quantum fluctuations}
\label{sec:coh-spinflip}
We now investigate the effect of quantum fluctuations in the form of non-zero coherences $\langle a^\dag_{l\uparrow} a_{l\downarrow}\rangle\ne0$, as they are  induced by the transverse magnetic field.  In Fig.~\ref{pd_H:ges}, we plot phase diagrams for a small system of 20 lattice sites for different values of $\omega$, both in the $T-D$ and in the $T-\rho_0$ plane. Note that the range of plotted temperatures is different in each plot. 

From the $T-D$ phase diagrams (upper panels of Fig.~\ref{pd_H:ges}), we can, first of all, clearly see that both the flocking and the aster phase are suppressed with increasing $\omega$, as expected \cite{quantum_flocking}. Namely, the corresponding transitions are shifted to lower temperatures. This effect is more pronounced for smaller hopping amplitudes $D$, leading to a non-monotonous behaviour of the critical temperature at which flocking occurs as a function of $D$. 

In order to investigate the competition between the suppression of collective effects by quantum fluctuations on the one hand and their enhancement by bosonic quantum statistics on the other, it is instructive to consider the density dependence of the critical transition temperature for flocking and aster formation and how it depends on $\omega$ (lower panels in Fig.~\ref{pd_H:ges}). As before, we can see that the critical temperatures for flocking and aster formation increase with the density. This can be explained by the fact that spin rotation induced by the magnetic field does not depend on the density (the rotation frequency is always $\omega$), whereas the dissipative processes responsible for the collective behaviour are linearly enhanced with the density via bosonic enhancement. 

\section{Conclusion}
\label{sec:conclusions}
We have proposed and investigated a quantum version of the active Ising model, which is based on ideal bosons with two relevant internal ``spin'' states.  The model shows all three phases known from the classical active Ising model in 1D: a disordered phase at high temperatures, a flocking phase at lower temperature, as well as a phase, where asters are formed. 
Compared to the classical model, we find a strongly increased stability of both collective behaviours, the formation of flocks and asters, as it is indicated  by markedly increased critical temperatures for both the flocking and the aster phase. This is a quantum statistical effect that can be attributed to the tendency of indistinguishable bosons to populate the same quantum state, as it is reflected in bosonic enhancement factors in transition amplitudes. 

This quantum statistical effect is not present in another recently introduced quantum generalization of the active Ising model, based on hard-core bosons \cite{quantum_flocking}.  In this model also no aster phases were found or discussed \cite{quantum_flocking}. This is likely a consequence of the hard-core constraint. The authors showed however, that the quantum fluctuations induced by a transverse magnetic field has a destabilizing effect on flocking, as it counteracts alignment.
Similarly, here we have studied the competition of quantum statistical enhancement on the one hand and its suppression via quantum fluctuations on the other, and found that for large densities, the quantum statistical enhancement of collective behavior dominates over its suppression via quantum fluctuations.

Our results open interesting directions for future research. One of them concerns (concepts for) the realization of active quantum systems, such as quantum active-Ising-type models, in quantum simulators via reservoir engineering and/or feedback control. Another one is to study the impact of quantum effects also on known non-equilibrium phases of classical active matter other than flocking and aster formation. This includes, for instance, flocking in higher dimensional systems \cite{Solon2013,ising3,scandolo_active_2023}, motility induced phase separation \cite{Cates2015,Bialke2013} or the appearance of time-dependent chiral states in non-reciprocal polar active systems \cite{Fruchart2021,Kreienkamp2025}. However, even more fascinating would be the discovery of unqiue quantum non-equilibrium phases, without classical counterpart. Finally, also the connection between quantum flocks and asters to non-equilibrium Bose-Einstein condensation \cite{Vorberg2013} is an interesting question to explore in future work. 

\begin{acknowledgements}
We acknowledge fruitful discussions with Hendrik Weimer, Markus Heyl and Michael te Vrugt.
    AE and AS acknowledge support by the Deutsche Forschungsgemeinschaft (DFG, German Research Foundation) via the Research Unit FOR 5688 (Project No. 521530974).
\end{acknowledgements}

\appendix
\crefalias{section}{appendix}
\section{Higher moments in kinetic equation for ideal bosons}\label{appendix_a}
In the dynamics without an external field, coherences, if not present initially, remain zero.
By investigating the expectation value of a general exponential function with number operators $n$ and $m$, we can rewrite
\begin{align}
    \langle e^{ n}\rangle \langle e^{ - m }\rangle = \sum_{k=0}^{\infty}\frac{\big\langle n^{k}\big\rangle}{k !} \sum_{l=0}^{\infty}\frac{\big\langle {(-m)}^{l}\big\rangle}{l !}\label{e-taylor}
\end{align}
Since higher momenta $\big\langle n^{k}\big\rangle$ cannot be implemented directly, we calculate it using Wick's Theorem. First we bring normal order
\begin{align}
    \big\langle n^{k}\big\rangle = \big\langle {(a^\dagger a)}^{k}\big\rangle = \sum_{j=1}^{k} \begin{Bmatrix}k \\ j\end{Bmatrix} \Big\langle{(a^\dagger)}^j a^j \Big\rangle \label{normal_order}
\end{align}
 with
${\{k,j\}}^T$ denoting the Stirling numbers of second kind.
After achieving a normally ordered expectation value, Wick's theorem can be applied as
\begin{align}
    \Big\langle {(a^\dagger)}^j a^j \Big\rangle = j! \langle n \rangle^j \label{wick}
\end{align}
to get the full identity 
\begin{align}
    \big\langle n^{k}\big\rangle = \sum_{j=1}^{k} \begin{Bmatrix}k \\ j\end{Bmatrix} j!\langle n \rangle^j.\label{n^k}
\end{align}
To express an exponential function we can then combine \ref{e-taylor} and \ref{n^k} to define the general expressions
\begin{align}
    \big\langle e^n \big\rangle &= 1+\sum_{k=1}^{\infty}\frac{1}{k!} \left(\sum_{q=1}^{k} \begin{Bmatrix}k \\ q \end{Bmatrix} q!\langle n \rangle^q \right) \label{en}\\
    \big\langle n e^n \big\rangle &= \langle n \rangle +\sum_{k=1}^{\infty}\frac{1}{k!} \left(\sum_{q=1}^{k+1} \begin{Bmatrix}k+1 \\ q\end{Bmatrix} q!\langle n \rangle^q \right) \label{nen}
\end{align}
Note that since ${\{0,q\}}^T$ is not defined, the first order term of the Taylor expansion of the exponential function is added by hand.
With these simple considerations, we have derived an equation that is limited in its accuracy only by the order up until which the exponential function is expanded. 

In the case of finite coherences, no simplifications can be made. Therefore, the expressions in \cref{D_12} and \cref{D_3456} are used.

\section{External Hamiltonian}\label{appendix_b}
To include quantum coherences into the master equation, in general the term $[H,\rho(t)]$ has to be included. 
Introducing a basic magnetic field Hamiltonian with frequency $\omega$ 
\begin{align}
    H&=\hbar \omega \left(a^\dagger_{k\uparrow} a_{k\downarrow} + a^\dagger_{k\downarrow} a_{k\uparrow} \right),
\end{align}
we calculate the expectation value of the commutator
\begin{align}
    \text{tr}& \left( n_{k \uparrow}[H,\rho(t)]\right) \notag\\
    =& \hbar \omega \,\text{tr}\left( a_{k \uparrow}^\dagger a_{k \uparrow} [a^\dagger_{k \uparrow} a_{k\downarrow} + a^\dagger_{k\downarrow} a_{k\uparrow},\rho(t)]\right)\notag\\
    =& \hbar \omega \,\text{tr} \Biggl( \left[a_{k \uparrow}^\dagger a_{k \uparrow} a^\dagger_{k\uparrow} a_{k\downarrow} \rho(t)+ a_{k \uparrow}^\dagger a_{k \uparrow} a^\dagger_{k\downarrow} a_{k\uparrow}\rho(t)\right]\notag\\
    &- \left[a^\dagger_{k\uparrow} a_{k\downarrow} a_{k \uparrow}^\dagger a_{k \uparrow} \rho(t) + a^\dagger_{k\downarrow} a_{k\uparrow} a_{k \uparrow}^\dagger a_{k \uparrow} \rho(t)\right]\Biggr).
    \label{H_1}
\end{align}
To obtain \cref{H_1}, the invariance of the trace under circular permutations was employed. 
Using the commutator $[a_m , a^\dagger_n]=\delta_{mn}$, the expectation values can be brought into normal ordering.
Combining the summands in \cref{H_1}, the resulting expression is
\begin{align}
    \text{tr}\left( n_{k \uparrow}[H,\rho(t)]\right) &= \text{tr} \left(a_{k \uparrow}^\dagger a_{k\downarrow} \rho(t) - a^\dagger_{k\downarrow} a_{k \uparrow} \rho(t)\right) \notag \\
    &= \langle a_{k \uparrow}^\dagger a_{k\downarrow} \rangle - \langle a^\dagger_{k\downarrow} a_{k \uparrow} \rangle.
\end{align}
This gives
\begin{align}
    \timederiv \langle n_{k\uparrow} \rangle_B &=-i\omega \left(\langle a_{k \uparrow}^\dagger a_{k\downarrow} \rangle - \langle a^\dagger_{k\downarrow} a_{k \uparrow} \rangle \right).
\end{align}
Analogously one obtains
\begin{align}
    \timederiv \langle n_{k\downarrow} \rangle_B &=-i\omega \left(\langle a_{k \downarrow}^\dagger a_{k\uparrow} \rangle - \langle a^\dagger_{k\uparrow} a_{k \downarrow} \rangle \right). 
\end{align}

The same derivation of the commutator has to be performed for the quantum coherences $\langle a_{k \uparrow/\downarrow}^\dagger a_{k \downarrow/\uparrow}\rangle$.
This is done in the exact same manner.
The resulting expectation value of the commutator is
\begin{align}
    \timederiv \langle a^\dagger_{k\uparrow } a_{k\downarrow}\rangle_B &=-i\omega \left(\langle n_{k \uparrow} \rangle - \langle n_{k\downarrow} \rangle \right),
\end{align}
and analogously 
\begin{align}
    \timederiv \langle a^\dagger_{k\downarrow} a_{k\uparrow}\rangle_B &=-i\omega \left(\langle n_{k\downarrow} \rangle - \langle n_{k \uparrow} \rangle \right)
\end{align}
for the opposite spins.
To implement the coherences numerically, another kinetic equation due to the Lindbladian has to be derived, which governs the dynamics and time evolution of the quantum coherences.\\

\section{Kinetic equations for coherences}
\label{app:ke-coh}
The dynamics of quantum coherences such as $\langle a_{k\uparrow}^\dagger a_{k\downarrow}\rangle$ can be described similarly to the kinetic equations for the occupation number operators. To do so, we begin by calculating the contribution from
\begin{align}
    \timederiv\langle & a^\dagger_{k\uparrow}a_{k\downarrow}\rangle_{1}\notag\\
    = &\text{tr} \left( a^\dagger_{k \uparrow} a_{k \downarrow} \left[ \gamma \sum_{l} \mathcal{D} \left[ a_{l \uparrow}^\dagger a_{l \downarrow} e^{\beta^*(n_{l \uparrow} - n_{l \downarrow})}\right]\superopbr{\rho}\right] \right)\notag \\
    \begin{split}
    =&\gamma \sum_{l} \text{tr} \Biggl( e^{\beta^*(n_{l \uparrow} - n_{l \downarrow})} a_{l \downarrow}^\dagger a_{l \uparrow} a^\dagger_{k \uparrow} a_{k \downarrow} a_{l \uparrow}^\dagger a_{l \downarrow} e^{\beta^*(n_{l \uparrow} - n_{l \downarrow})}\rho \\
    &- \frac{1}{2} a^\dagger_{k \uparrow} a_{k \downarrow} e^{\beta^*(n_{l \uparrow} - n_{l \downarrow})} a_{l \downarrow}^\dagger a_{l \uparrow} a_{l \uparrow}^\dagger a_{l \downarrow} e^{\beta^*(n_{l \uparrow} - n_{l \downarrow})} \rho \\
    &- \frac{1}{2} e^{\beta^*(n_{l \uparrow} - n_{l \downarrow})} a_{l \downarrow}^\dagger a_{l \uparrow} a_{l \uparrow}^\dagger a_{l \downarrow} e^{\beta^*(n_{l \uparrow} - n_{l \downarrow})} a^\dagger_{k \uparrow} a_{k \downarrow} \rho \Biggr)\label{D_11}
    \end{split}
\end{align}
To arrive at \cref{D_11}, we have used the invariance of the trace under circular permutations. 
With the density matrix $\rho$ at the end of each summand, the rest of the operators will be brought into the same order to simplify the expression. 
One useful relation for this is
\begin{align}
    a^\dagger_{k \uparrow} a_{k \downarrow} e^{\beta^*(n_{l \uparrow} - n_{l \downarrow})} &= e^{\beta^*(n_{l \uparrow} - n_{l \downarrow})} a^\dagger_{k \uparrow} a_{k \downarrow} e^{\beta^*}\delta_{kl},
\end{align}
which can be derived using the Baker-Campbell-Hausdorff formula.
Employing this, the second and third summand can be brought into the same form as the first.
Combining them with \cref{D_11}, we get the final equation \cref{unwick_D1} and analogously \cref{unwick_D2},
\begin{widetext}
\begin{align}
\begin{split}
    \timederiv\langle a^\dagger_{k\uparrow}a_{k\downarrow}\rangle_{1}= &\gamma \Biggl\langle e^{\beta^*(n_{k \uparrow} - n_{k \downarrow})} a_{k \downarrow}^\dagger a_{k \uparrow} a^\dagger_{k \uparrow} a_{k \downarrow} a_{k \uparrow}^\dagger a_{k \downarrow} e^{\beta^*(n_{k \uparrow} - n_{k \downarrow})}\left(1-\frac{e^{\beta^*}}{2}-\frac{e^{-\beta^*}}{2}\right)\\
    &+e^{\beta^*(n_{k \uparrow} - n_{k \downarrow})} a_{k \downarrow} a_{k \downarrow}^\dagger a_{k \uparrow}^\dagger a_{k \downarrow} e^{\beta^*(n_{k \uparrow} - n_{k \downarrow})} \frac{e^{\beta^*}}{2} -e^{\beta^*(n_{k \uparrow} - n_{k \downarrow})} a^\dagger_{k \uparrow} a_{k \uparrow} a_{k \uparrow}^\dagger a_{k \downarrow} e^{\beta^*(n_{k \uparrow} - n_{k \downarrow})}\frac{e^{\beta^*}}{2} \Biggr\rangle \label{unwick_D1}\\
\end{split}
\end{align}
\begin{align}
\begin{split}
    \timederiv\langle a^\dagger_{k\downarrow}a_{k\uparrow}\rangle_{2} =&\gamma \Biggl\langle  e^{-\beta^*(n_{k \uparrow} - n_{k \downarrow})} a_{k \uparrow}^\dagger a_{k \downarrow} a^\dagger_{k \uparrow} a_{k \downarrow} a_{k \downarrow}^\dagger a_{k \uparrow} e^{-\beta^*(n_{k \uparrow} - n_{k \downarrow})} \left(1-\frac{e^{\beta^*}}{2}-\frac{e^{-\beta^*}}{2}\right)\\
    &-e^{-\beta^*(n_{k \uparrow} - n_{k \downarrow})} a_{k \uparrow}^\dagger a_{k \downarrow} a_{k \downarrow}^\dagger a_{k \downarrow} e^{-\beta^*(n_{k \uparrow} - n_{k \downarrow})} \frac{e^{\beta^*}}{2} +e^{-\beta^*(n_{k \uparrow} - n_{k \downarrow})} a_{k \uparrow}^\dagger a_{k \downarrow} a_{k \uparrow} a^\dagger_{k \uparrow} e^{-\beta^*(n_{k \uparrow} - n_{k \downarrow})} \frac{e^{\beta^*}}{2} \Biggr\rangle \label{unwick_D2}
\end{split}
\end{align}
\end{widetext}
We combine the first summands of both $\langle  a^\dagger_{k\uparrow}a_{k\downarrow}\rangle_{1}$ and $\langle a^\dagger_{k\uparrow}a_{k\downarrow}\rangle_{2}$  and employ the Taylor expansion $e^{\pm\beta^*(n_{k \uparrow} - n_{k \downarrow})}\approx 1 \pm \beta^*(n_{k \uparrow} - n_{k \downarrow})$. Here, all summands containing ${\beta^*}^2$ are set to zero in order to correctly adapt the Taylor expansion into the final equation. 
\begin{align}
 \timederiv \langle a^\dagger_{k \uparrow} a_{k\downarrow}\rangle_\mathrm{F} 
    =&\gamma
    \Bigl\langle \{a_{k \uparrow}^\dagger a_{k \downarrow} n_{k \downarrow} , \beta^*(n_{k \uparrow} - n_{k \downarrow})\} \Bigr\rangle \notag\\
    &+\gamma\Bigl\langle \{n_{k \uparrow} a^\dagger_{k \uparrow} a_{k \downarrow} , \beta^*(n_{k \downarrow} - n_{k \uparrow})\} \Bigr\rangle. \label{D_12_0}
\end{align}
We note that the expectation values still contain a product of many operators.
Evaluating the expression in \cref{D_12_0} under a mean-field approximation is  lengthy. We expand the anticommutators, put the operators into normal order and apply Wick's theorem. This brings us to the final flipping equation found in \cref{D_12}.\\

\textit{Hopping Equation:}
Additionally, the hopping terms in the kinetic equation need to be calculated. In the following, this will be carried out as an example for
\begin{align}
    \timederiv\langle a^\dagger_{k\uparrow}a_{k\downarrow}\rangle_{3}=&D (1+ \varepsilon) \text{tr} \left\lbrace a^\dagger_{k \uparrow} a_{k \downarrow}  \sum_{l}  \left(  \mathcal{D}\left[ a_{l+1 \uparrow}^\dagger a_{l \uparrow}\right]\superopbr{\rho}\right) \right\rbrace \notag\\
    =&D (1+\varepsilon) \sum_l  \text{tr} \Biggl(a^\dagger_{k \uparrow} a_{k \downarrow} a_{l+1 \uparrow}^\dagger a_{l \uparrow} \rho a_{l \uparrow}^\dagger a_{l+1 \uparrow} \notag\\
    &-\frac{1}{2} \Big(a^\dagger_{k \uparrow} a_{k \downarrow} a_{l \uparrow}^\dagger a_{l+1 \uparrow} a_{l+1 \uparrow}^\dagger a_{l \uparrow} \rho \notag\\
    &+a^\dagger_{k \uparrow} a_{k \downarrow} \rho a_{l \uparrow}^\dagger a_{l+1 \uparrow} a_{l+1 \uparrow}^\dagger a_{l \uparrow}\Big)\Biggr)\label{D_31}
\end{align}
Using the invariance in the trace 
as well as the commutator $[a^\dagger_i a_j, a^\dagger_k]=a_i^\dagger \delta_{jk}$ or $[a^\dagger_k, a^\dagger_i a_j]=-a_i^\dagger \delta_{jk}$ respectively, we can rewrite the first summand in \cref{D_31}.
\begin{align}
\begin{split}
    &a^\dagger_{k \uparrow} a_{k \downarrow} a_{l \uparrow}^\dagger a_{l+1 \uparrow} a_{l+1 \uparrow}^\dagger a_{l \uparrow}  \\
    &= a_{l \uparrow}^\dagger a_{l+1 \uparrow} a^\dagger_{k \uparrow} a_{k \downarrow} a_{l+1 \uparrow}^\dagger a_{l \uparrow} 
    - \delta_{l+1 k}a^\dagger_{\uparrow l} a_{k \downarrow} a_{l+1 \uparrow}^\dagger a_{l \uparrow}, 
\end{split}
\end{align}
Doing the same for the second and third summand shows,
that the two terms of the resulting anticommutator cancel out with the leading summand, leaving us with 
\begin{align}
\begin{split}
    \timederiv\langle a^\dagger_{k\uparrow}a_{k\downarrow}\rangle_{3}=& D\frac{(1+\varepsilon)}{2} \Bigl\langle a_{k \uparrow}^\dagger a^\dagger_{k-1 \uparrow} a_{k-1 \uparrow} a_{k \downarrow} \\&- a_{k \uparrow}^\dagger a_{k+1 \uparrow}^\dagger a_{k+1 \uparrow} a_{k \downarrow} - a_{k \uparrow}^\dagger a_{k \downarrow}\Bigr\rangle\label{D_33}
    \end{split}
\end{align}
Similar results can be found for $\langle a^\dagger_{k\uparrow}a_{k\downarrow}\rangle_{4-6}$
\begin{align}
\begin{split}
    \timederiv\langle a^\dagger_{k\uparrow}a_{k\downarrow}\rangle_{4}= &\text{tr} \left( a^\dagger_{k\uparrow } a_{k \downarrow } \left[ D \sum_{l} (1+\varepsilon) \mathcal{D} \left[ a_{l-1 \downarrow}^\dagger a_{l \downarrow}\right]\superopbr{\rho}\right] \right) \\
    =&D\frac{(1+\varepsilon)}{2} \Bigl\langle a^\dagger_{k \uparrow } a_{k+1 \downarrow}^\dagger a_{k+1 \downarrow} a_{k \downarrow} \\&- a^\dagger_{k \uparrow } a_{k-1 \downarrow}^\dagger a_{k-1 \downarrow} a_{k \downarrow} - a^\dagger_{k \uparrow } a_{k \downarrow} \Bigr\rangle\label{D_44}
\end{split}
\end{align}
\begin{align}
\begin{split}
    \timederiv\langle a^\dagger_{k\uparrow}a_{k\downarrow}\rangle_{5}=& \text{tr} \left( a^\dagger_{k \uparrow} a_{k \downarrow} \left[ D \sum_{l} (1-\varepsilon) \mathcal{D} \left[ a_{l-1 \uparrow}^\dagger a_{l \uparrow}\right]\superopbr{\rho}\right] \right) \\
    =&D\frac{(1-\varepsilon)}{2} \Bigl\langle a_{k \uparrow}^\dagger a^\dagger_{k+1 \uparrow} a_{k+1 \uparrow} a_{k \downarrow}\\&- a_{k \uparrow}^\dagger a_{k-1 \uparrow}^\dagger a_{k-1 \uparrow} a_{k \downarrow} - a_{k \uparrow}^\dagger a_{k \downarrow}\Bigr\rangle\label{D_55}
\end{split}
\end{align}
\begin{align}
\begin{split}
    \timederiv\langle a^\dagger_{k\uparrow}a_{k\downarrow}\rangle_{6}=& \text{tr} \left( a^\dagger_{ k\uparrow} a_{ k\downarrow} \left[ D \sum_{l} (1-\varepsilon) \mathcal{D}\left[ a_{l+1 \downarrow}^\dagger a_{l \downarrow}\right]\superopbr{\rho}\right] \right) \\
    =&D\frac{(1-\varepsilon)}{2} \Bigl\langle a^\dagger_{k\uparrow } a_{k-1 \downarrow}^\dagger  a_{k-1 \downarrow} a_{k \downarrow} \\&- a^\dagger_{k \uparrow } a_{k+1 \downarrow}^\dagger a_{k+1 \downarrow} a_{k \downarrow} - a^\dagger_{k \uparrow } a_{k \downarrow} \Bigr\rangle \label{D_66}
\end{split}
\end{align}
 Combining them, the resulting dissipative hopping equation is
\begin{align}
\begin{split}
    \timederiv \langle a^\dagger_{k\uparrow } a_{ k \downarrow}\rangle_\mathrm{H} =& D \varepsilon \Bigl\langle  a_{k \uparrow}^\dagger a_{k \downarrow} n_{k-1 \uparrow} + a^\dagger_{k\uparrow } a_{k \downarrow} n_{k+1 \downarrow} \\&- a_{k \uparrow}^\dagger a_{k \downarrow} n_{k+1 \uparrow}- a^\dagger_{k\uparrow } a_{k \downarrow} n_{k-1 \downarrow} - \frac{2}{\varepsilon} a^\dagger_{k\uparrow } a_{k \downarrow}\Bigr\rangle\label{D_3456_0}
\end{split}
\end{align}
Note that \cref{D_3456_0} can be trivially decomposed in expectation values, since we neglect coherences on neighbouring sites such as $\langle a^\dagger_{k \uparrow} a_{k+1 \downarrow}\rangle$.

\section{Estimating the critical temperature for flocking}\label{appendix_d}
In the absence of an external field, equations of motion for the systems magnetisation $m_k = n_{k\uparrow} - n_{k\downarrow}$ and density $\rho_k = n_{k\uparrow} + n_{k\downarrow}$ can be derived by inserting these definitions into \cref{eq:n-mean-compl} for $\timederiv\langle n_{k\uparrow}\rangle$ and $\timederiv\langle n_{k\downarrow}\rangle$. 
For a steady state in the flocking phase, both the magnetization as well as the density are homogeneous with global magnetization 
$\langle m \rangle = \langle n_{\uparrow}\rangle - \langle n_{\downarrow}\rangle$ and global density $\rho_0 = \langle n_{\uparrow}\rangle + \langle n_{\downarrow}\rangle$. Then, the mean-field equations simplify to
\begin{align}
    \timederiv \rho_0=0
    \label{eq:dt rho}
\end{align}
and
\begin{align}
\begin{split}
    \timederiv \langle m\rangle =& 2\gamma   \langle(1+n_{\uparrow}) e^{ \beta^* n_{\uparrow}}  \rangle \langle n_{\downarrow} e^{ -\beta^* n_{\downarrow}}\rangle\\
 &-2\gamma \langle(1+n_{\downarrow}) e^{ \beta^* n_{\downarrow}}  \rangle \langle n_{\uparrow} e^{ -\beta^* n_{\uparrow}}\rangle\overset{!}{=}0.
\end{split}
    \label{eq;dt m}
\end{align}
We investigate Eq. (\ref{eq;dt m}) further by first expanding the exponentials as discussed in Appendix \ref{appendix_a}, 
\begin{align}
\begin{split}
    \timederiv\langle m\rangle = &2\gamma \Big(\langle m\rangle (2\beta -1)   
     + \beta \frac{\langle m\rangle}{\rho_0}(1+4 \langle n_\uparrow \rangle\langle n_\downarrow\rangle)\Big)\overset{!}{=}0,
\end{split}
\end{align}
and then rewriting $\langle n_{\uparrow}\rangle$ and $\langle n_{\downarrow}\rangle$ in terms of $\langle m\rangle$ and $\rho_0$, where $\langle n_{\uparrow}\rangle =({\rho_0+\langle m\rangle})/{2}$, $\langle n_{\downarrow}\rangle = ({\rho_0-\langle m\rangle})/{2}$ and $\langle n_{\uparrow}\rangle\langle n_{\downarrow}\rangle = ({\rho_0^2-\langle m\rangle^2})/{4}$.
We find 
\begin{align}
\begin{split}
    \timederiv \langle m\rangle =& 2\gamma\Big( \langle m\rangle(2\beta -1)  + \beta\frac{\langle m\rangle}{\rho_0}(1+\rho_0^2-\langle m\rangle^2)\Big)\overset{!}{=}0.
\intertext{This can be rewritten as}
   0 =& \langle m \rangle^3 - \langle m\rangle \Big( (\rho_0 + 1 )^2 - \frac{\rho_0}{\beta} \Big),
    \label{eq:m}
\end{split}
\end{align}
which is a cubic equation in $\langle m \rangle$ of the form $m^3-am=0$ with the solutions $\langle m_1\rangle = 0$ and  $\langle m_{2/3}\rangle = \pm\sqrt{a}$. Since $\langle m\rangle\in\mathbb{R}$, solutions with flocking order, i.e.~$\langle m \rangle \neq 0$, can only exist if
\begin{align}
\begin{split}
    (\rho_0 + 1 )^2 &> \frac{\rho_0}{\beta} \quad 
     \Leftrightarrow \quad T < \frac{(\rho_0 + 1 )^2}{\rho_0}.
\end{split}
\end{align}
\\
\\
\\

\bibliography{apssamp}

\end{document}